\documentclass[letterpaper,aps,prd,twocolumn,tightenlines,preprintnumbers,nofootinbib,showkeys,superscriptaddress,longbibliography
]{revtex4-1}

\pdfoutput=1 % if your are submitting a pdflatex (i.e. if you have
\usepackage[T1]{fontenc}

\usepackage{fullpage}
\usepackage{amsfonts}
\usepackage{amsmath}
\usepackage{slashed}
\usepackage{amssymb}
\usepackage{graphicx}
\usepackage{epic}
\usepackage{eepic}
\usepackage{epsfig}
\usepackage{latexsym}
\usepackage[dvipsnames]{xcolor}
\usepackage{float}
\usepackage{multirow}
\usepackage[breaklinks]{hyperref}
\usepackage{enumitem}
\usepackage[caption=false]{subfig}
\usepackage{natbib}
\usepackage{relsize}
\usepackage[left=2cm,right=2cm,top=2cm,bottom=2cm]{geometry}
\usepackage{makecell}
\usepackage{soul}
\usepackage{bm}
\usepackage{subfig}
\usepackage{shorthand}
\usepackage{mciteplus}

\hypersetup{colorlinks=true,citecolor=blue,linkcolor=blue,urlcolor=NavyBlue}

\begin{document}
%\relscale{1.05}
\linespread{1.02}

%\title{$\mathbf{B}$-flavour and $\mathbf{(g-2)_\mu}$ anomalies in a 2HDM with leptoquarks from $\mathbf{SO(10)}$ grand unification}

\title{Economical model for $B$-flavour and $a_\mu$ anomalies from $\mathbf{SO(10)}$ grand unification}

% $\mathbf{\textit{S}_1}$ 

\author{Ufuk Aydemir}
\email{uaydemir@ihep.ac.cn}
\affiliation{Department of Physics, Middle East Technical University, Ankara 06800, T\"urkiye}
\affiliation{Institute of High Energy Physics, Chinese Academy of Sciences, Beijing 100049, P. R. China}

\author{Tanumoy Mandal}
\email{tanumoy@iisertvm.ac.in}
\affiliation{Indian Institute of Science Education and Research Thiruvananthapuram, Vithura, Kerala, 695 551, India}

\author{Subhadip Mitra}
\email{subhadip.mitra@iiit.ac.in}
\affiliation{Center for Computational Natural Sciences and Bioinformatics, International Institute of Information Technology, Hyderabad 500 032, India}

\author{Shoaib Munir}
\email{smunir@eaifr.org}
\affiliation{East African Institute for Fundamental Research (ICTP-EAIFR), University of Rwanda, Kigali, Rwanda}
\affiliation{Department of Physics, Faculty of Natural Sciences and Mathematics, St. Olaf College, Northfield, MN 55057, USA}

\date{\today}

\begin{abstract}

We investigate an $\mathrm{SO}(10)$ grand unification scenario where the complex 10-dimensional scalar multiplet, containing the Standard Model (SM) Higgs boson, resides at the TeV scale altogether. The resulting low-energy model is a 2-Higgs-doublet model augmented with two $S_1$-type leptoquarks. The gauge-coupling unification is achieved with only one intermediate scale at which the Pati-Salam gauge group is broken down to the SM. Proton stability is ensured by a discrete symmetry, leftover from the breaking of the U(1)$_{\mathrm{PQ}}$ global symmetry. The axion corresponding to this broken U(1)$_{\mathrm{PQ}}$ provides a solution to the strong CP problem, and also serves as an important dark matter candidate. We discuss how the simultaneous explanation for the $R_{D^{(*)}}$ and $a_\mu$ anomalies comes about in the model, and investigate some of its phenomenological implications. 

\end{abstract}

\keywords{$B$-decay anomalies, scalar leptoquark, $\mathrm{SO}(10)$ grand unification, LHC, 2HDM, axions, Peccei-Quinn symmetry, strong CP problem, dark matter} 

\maketitle

\section{Introduction} \label{sec:intro}

During the past few years, several disagreements between particle physics experiments and the SM predictions of the rare $B$-decays have been reported by the BaBar~\cite{BaBar:2012obs,BaBar:2013mob}, LHCb~\cite{LHCb:2014vgu,LHCb:2017avl,LHCb:2015gmp,LHCb:2017smo,LHCb:2017rln}, and Belle~\cite{Belle:2016dyj,Belle:2017ilt} collaborations. A pair of such anomalous measurements that has persisted (not without significant fluctuations, though) corresponds to the $R_{D^{(*)}}$ observables, defined as
\be
R_{D^{(*)}} = \dfrac{\mathrm{BR}(B\rightarrow D^{(*)}\tau\nu)}{\mathrm{BR}(B\rightarrow D^{(*)}\ell\nu)}\,,
\ee
where $\ell=\{e,\,\mu\}$, and $\mathrm{BR}$ stands for branching ratio. The latest averaged experimental values of these observables read \cite{Hflav-web}
\ba
%R_D  &=  0.340\pm 0.027~({\rm stat})\pm 0.013~({\rm syst})\;,\nn\\
%R_{D^*} &=  0.295\pm 0.011~({\rm stat})\pm 0.008~({\rm syst})\;,
%%%%%%%%%%%%%%%% 2023 average %%%%%%%%%%%%%%%%%%%%%%%%%%%
%R_D  &=  0.356 \pm 0.029\;,\nn\\ % 0.327-0.385
%R_{D^*} &= 0.283 \pm 0.013   % 0.270-0.296
%%%%%%%%%%%%%%%% 2024 post-Moriond average %%%%%%%%%%%%%%
R_D  &=&  0.342 \pm 0.026\;,\nn\\ % 0.316-0.368
R_{D^*} &=& 0.287 \pm 0.012  % 0.275-0.299
\ea
which exceed the averaged SM predictions, $R_D^{\rm SM} = 0.298$ and $R_{D^*}^{\rm SM} = 0.254$, estimated by the Heavy Flavor Averaging Group~\cite{HFLAV:2022esi} by more than 3$\sigma$ (taking into account the $R_D$-$R_{D^*}$ correlation) \cite{Hflav-web}. Another persistent collider anomaly is the measurement of the magnetic moment of the muon, $a_\mu$. Its current BNL \cite{Muong-2:2006rrc} and FNAL~\cite{Muong-2:2023cdq} experimental average, $\{116\,592\,059 \pm 22\}\times 10^{-11}$, deviates from $a_\mu^{\rm SM} = \{116\,591\,810 \pm 43\}\times 10^{-11}$~\cite{Czarnecki:2002nt,Melnikov:2003xd,Aoyama:2012wk,Gnendiger:2013pva,Kurz:2014wya,Colangelo:2014qya,Colangelo:2017fiz,Davier:2017zfy,Masjuan:2017tvw,Keshavarzi:2018mgv,Hoferichter:2018kwz,Colangelo:2018mtw,Gerardin:2019vio,Hoferichter:2019mqg,Davier:2019can,Bijnens:2019ghy,Colangelo:2019uex,Blum:2019ugy,Keshavarzi:2019abf,Aoyama:2019ryr,Aoyama:2020ynm} by 5.1$\sigma$, so that
\be
\Delta a_\mu = \{24.9 \pm 4.8\}\times 10^{-10}\;.%2304.00038
\ee
%new average: a_\mu = (116\,592\,059 \pm 22)\times 10^{-11}, $\Delta a_\mu = (24.9 \pm 4.8)\times 10^{-10}$~\cite{2304.00038} (10/08/23 FNAL announcement, $5.1\sigma$)

There are many proposed new physics scenarios that could explain these anomalies, mostly introducing particles on an ad-hoc basis. However, a compelling rationale for the consistency of TeV-scale particles with the existing frameworks of ultraviolet (UV)-completion is mostly overlooked. While no such motivation is a requisite in a bottom-up approach, it is nevertheless appealing to search for a UV-compatible picture based on our current understanding of the SM. Motivated by this, we investigate here a low-energy model originating from a grand unified theory (GUT).

The existence of some particles in a GUT framework at the TeV scale could be approached in the context of the splitting of the parent multiplet containing the SM Higgs field into two (or more) mass scales. A large mass-splitting is a well-known problem in supersymmetric and non-supersymmetric GUTs both, and can, quite obviously, be circumvented (at least to some extent) if the companions of the doublet Higgs field in the $\mathrm{SO}(10)$ multiplet also lie at the TeV scale. The null results from the new particle searches at the large hadron collider (LHC), and at the same time the anticipation about its upcoming extended run(s), make the relatively small and economical TeV-scale multiplet, such as a 10-dimensional scalar representation $\mathbf{10}_H$, a rather appealing prospect. This representation ordinarily implies a real multiplet which can contain only the electroweak (EW) Higgs doublet along with a $S_1$-type scalar leptoquark, which we generically denote by $\widetilde{L}$ henceforth. Such a leptoquark has been extensively studied in the literature~\cite{Sakaki:2013bfa,Dumont:2016xpj,Freytsis:2015qca,Bauer:2015knc,Das:2016vkr,Hiller:2016kry,ColuccioLeskow:2016dox,Crivellin:2017zlb,Cai:2017wry,Marzocca:2018wcf,Aydemir:2018cbb,Becirevic:2018afm,Angelescu:2018tyl,Heeck:2018ntp,Mandal:2018kau,Aydemir:2019ynb,Crivellin:2019dwb,Marzocca:2021azj,Bhaskar:2022vgk,Bhaskar:2024swq} as a plausible explanation for the $R_{D^{(*)}}$ and the $a_\mu$ anomalies, although it has now been realised that a single $\widetilde{L}$ cannot explain all these anomalies simultaneously~\cite{Angelescu:2018tyl}.\footnote{We note here that the 3.1$\sigma$ deviation previously reported by the LHCb collaboration \cite{LHCb:2019hip} in the measurement of the $R_{K^{(*)}} \equiv \dfrac{\mathrm{BR}(B\rightarrow K^{(*)}\mu^+\mu^-)}{\mathrm{BR}(B\rightarrow K^{(*)}e^+e^-)}$ observables from their SM prediction of $1.00\pm 0.01$ has recently been refuted \cite{LHCb:2022qnv}.}

Extension of the SM by a solo $\widetilde{L}$ resulting from a real $\mathrm{SO}(10)$ multiplet, $\mathbf{10}_H$, has been studied in Ref.~\cite{Aydemir:2019ynb} to address only the $R_{D^{(*)}}$ anomalies. The more popular scenario in the literature is the one with the complex(ified) version of the $\mathbf{10}_H$ multiplet~\cite{Bajc:2005zf,Babu:2015bna}, which has its advantages in the context of the fermion mass relations and requires less number of large GUT multiplets compared to the real case~\cite{Babu:2016bmy}.  The complex multiplet $\mathbf{10}_H$ yields two Higgs doublets along with two $\widetilde{L}$ fields, which are assumed to lie at the TeV-scale in our scenario. The fact that TeV-scale 2-Higgs doublet models (2HDMs) are themselves phenomenologically favorable for multiple reasons (see~\cite{Branco:2011iw} for a review) makes this scenario even more appealing to explore.

The TeV-scale particle content of the model does not lead to gauge-coupling unification at the purported GUT scale, where the masses of the rest of the particles are assumed to lie. Therefore, we follow the common route of inserting a single intermediate stage where the active gauge group is the Pati-Salam group, $\mathrm{SU}(4)_{\mathcal{C}}\otimes \mathrm{SU}(2)_L\otimes \mathrm{SU}(2)_R$, the breaking of which appears to be favored by various phenomenological bounds~\cite{Altarelli:2013aqa}. In our model, the $\mathrm{SO}(10)$ symmetry is thus broken at the unification scale $M_U$ into the Pati-Salam gauge group, which itself subsequently breaks into the SM gauge group at an intermediate energy scale $M_{\mathrm{PS}}$. 

Furthermore, a Peccei-Quinn global symmetry, U(1)$_{\mathrm{PQ}}$~\cite{Peccei:1977hh,Weinberg:1977ma,Wilczek:1977pj}, is introduced along with a singlet scalar, which provides the axion that resolves the strong CP problem and constitutes a dark matter (DM) candidate upon acquiring a vacuum expectation value (VEV)~\cite{Babu:1992ia,Bajc:2005zf,Babu:2015bna,Ernst:2018bib}. The couplings of the leptoquarks that can potentially lead to proton decay are forbidden by imposing a discrete symmetry, identified with the baryon number, as we will discuss in the next section.

A 2HDM, which the Higgs sector of our low-energy model resembles, can itself explain $\Delta a_\mu$, albeit in a very narrow parameter space region, due to the presence of a charged scalar, $H^\pm$~\cite{Broggio:2014mna,Wang:2014sda,Ilisie:2015tra,Omura:2015nja,Abe:2015oca,Chun:2015hsa,Chun:2015xfx,Han:2015yys,Chun:2016hzs,Cherchiglia:2016eui,Cherchiglia:2017uwv,Wang:2018hnw,Chun:2019oix,Iguro:2019sly,Chun:2019sjo,Jana:2020pxx,Li:2020dbg,Keung:2021rps,Ferreira:2021gke,Han:2021gfu,Chun:2021rtk,Jueid:2021avn,Athron:2021iuf,Hou:2021qmf,Dey:2021pyn,Hou:2021sfl}. It has also been studied as a prospect accommodating the $R_{D^{(*)}}$ anomalies~\cite{Athron:2021auq}, and found to serve the purpose only with additional structure~\cite{Crivellin:2015hha,Chen:2017eby,Li:2018rax,Ghosh:2020tfq,Chen:2021vzk} (again, in the bottom-up approach). In the model explored in this article, the synergy between the two copies of $\widetilde{L}$ augmenting the GUT-inspired 2HDM at the low energy provides a rather natural recipe for explaining these anomalies. Our analysis shows that even in the most minimal scenario of this model, they can indeed be successfully addressed simultaneously.

The rest of the article is organised as follows. In Sec. \ref{sec:model} we present at length the GUT framework of our interest, as well as the resulting low-energy model. In Sec. \ref{sec:pheno} we overview the model's parameter space of our interest. This is followed  in Sec. \ref{sec:numeric} by the details of the analysis of our low-energy model's predictions against the experimental data. In Sec. \ref{sec:results} we discuss some of the phenomenological implications of our model, and in Sec \ref{sec:Conc} we conclude our findings.

%%%%%%%%%%%%%%%%%%%%%%%%%%%%%%%%%%%%%%%%%%%%%%%%%%%%%%%%%%%%
\section{The $\mathbf{SO(10)}$ model}
\label{sec:model}
%%%%%%%%%%%%%%%%%%%%%%%%%%%%%%%%%%%%%%%%%%%%%%%%%%%%%%%%%
\subsection{The UV framework}
In the $\mathrm{SO}(10)$ framework, each family of the SM fermions, augmented by right-handed neutrinos, resides in a $\mathbf{16}$ spinor-representation. As for the scalar field content responsible for the Yukawa sector, a combination of $\mathbf{10}$, $\mathbf{120}$ and $\mathbf{126}$ representations must be selected, since
\begin{eqnarray}
\mathbf{16}\otimes \mathbf{16}=\mathbf{10}\oplus \mathbf{120}\oplus\mathbf{126}\;.
\end{eqnarray}
In this paper, for the UV part, we will adopt the common framework wherein the scalar sector consists of a complex(ified) $\mathbf{10}_{H}$ and a complex $\mathbf{126}_{H}$, which is known to yield a realistic Yukawa sector (e.g. fermion mass relations and mixing patterns) as well as a successful see-saw mechanism for the neutrino masses~\cite{Babu:1992ia,Bajc:2005zf,Babu:2015bna}.\footnote{A real $\mathbf{10}_{H}$, instead of the complex one, is also a phenomenologically viable option~\cite{Aydemir:2016qqj,Babu:2016bmy}, although the earlier discussions suggested otherwise~\cite{Bajc:2005zf}. In fact, we utilised in Ref.~\cite{Aydemir:2019ynb} a version of this SO(10) framework with a real $\mathbf{10}_{H}$ near the EW scale (and with a different $\mathrm{SO}(10)$ scalar content and Yukawa sector), which yields, besides the SM Higgs boson, a single TeV-scale $\widetilde{L}$ explaining the $R_{D^{(*)}}$ anomalies. The light complex $\mathbf{10}_{H}$, which is the case in this paper, yields a much richer structure, enabling us to additionally explain the $a_\mu$ anomaly.} 

In addition to these two multiplets, a $\mathbf{54}_{H}$ is utilised to break the $\mathrm{SO}(10)$ symmetry. Finally, we also have a scalar singlet $S$, which breaks the U(1)$_{\mathrm{PQ}}$ symmetry \cite{Peccei:1977hh,Weinberg:1977ma,Wilczek:1977pj}, imposed to resolve the strong CP problem, provides a potential DM candidate, and increases the predictivity of the theory by forbidding certain terms in the Lagrangian~\cite{Babu:1992ia,Bajc:2005zf,Babu:2015bna,Ernst:2018bib}. The last feature also leads to the so-called Type-II 2HDM along with two leptoquarks at the low energy (see Eq.~(\ref{eq:lagrangianLQ}) below), which is the main focus of this paper.

The Yukawa terms are given as 
\begin{eqnarray}
\label{eq:UVLagrangian}
\mathcal{L}_Y=\mathbf{16}_F\left(Y_{10}\mathbf{10}_H+Y_{126}\mathbf{\overline{126}}_H\right)\mathbf{16}_F
+ \mathrm{H.c.}\;,
\end{eqnarray}
where $Y_{10}$ and $Y_{126}$ are complex matrices, symmetric in the generation space. The $\mathbf{10}^*_H$ term is forbidden by the U(1)$_{\mathrm{PQ}}$ symmetry, the assigned charges of which are 
\begin{eqnarray}
&&\mathbf{16}_{F} \rightarrow e^{i \alpha} \mathbf{16}_{F}\;, \quad \mathbf{10}_{H} \rightarrow e^{-2 i \alpha} \mathbf{10}_{H}\;, \nonumber\\&& \mathbf{126}_{H} \rightarrow e^{2 i \alpha} \mathbf{126}_{H}\;, \quad S \rightarrow e^{-4 i \alpha} S\;.
\end{eqnarray}
\\
The VEV of the $\mathbf{126}_{H}$ field can only break a linear combination of U(1)$_{B-L}$ and U(1)$_{\mathrm{PQ}}$ (and hence the Pati-Salam group) into the SM group. Therefore, in order to obtain the Goldstone boson (axion), the singlet $S$ above has been introduced~\cite{Mohapatra:1982tc} to break the U(1)$_{\mathrm{PQ}}$ alone by acquiring a VEV $\left\langle S \right\rangle\equiv M_{\mathrm{PQ}}\simeq f_a$, where $f_a$ is the axion decay constant. This symmetry could also be broken by $\left\langle \mathbf{10}_H\right\rangle$, similarly to the original axion model~\cite{Peccei:1977hh,Weinberg:1977ma,Wilczek:1977pj},  but in that case one would obtain $M_{\mathrm{PQ}}=M_{\mathrm{EW}}$, which has long been ruled out~\cite{Kim:1986ax,Mimasu:2014nea} (see also Ref.~\cite{Marsh:2015xka} for a review). The scale $M_{\mathrm{PQ}}$ is independent of the unification scale $M_U$, and its value can be chosen in accordance with the axion phenomenology. In Ref.~\cite{Babu:2015bna} $M_{\mathrm{PQ}}\sim 10^{10}-10^{12}$ GeV was selected in a full numerical fit of the model parameters, which is consistent with the current constraints, and yields the axion mass $m_a \sim 10-200\;\mu \mathrm{eV}$, making it a suitable candidate for the DM of the Universe.

The Pati-Salam and SM decompositions of the $\mathbf{10}_H$  are given as
\begin{eqnarray}
\label{10H-decomp}
\mathbf{10}&=& \Phi\left(1,2,2\right)_{422} \oplus \Xi\left(6,1,1\right)_{422} \nonumber\\ \nonumber\\
&=&\phi_1\left(1,2,\frac{1}{2}\right)_{321}\oplus\; \;\phi_2\left(1,2,-\frac{1}{2}\right)_{321}\nn\\ &&\oplus\;\; \xi_1 \left(3,1,-\frac{1}{3}\right)_{321}\oplus \;\;\xi_2\left(\bar{3},1,\frac{1}{3}\right)_{321}\;,
\end{eqnarray}
where the subscripts denote the corresponding gauge group, and we have set $Q=I_3+Y$. $H_1$ and $H_2$ are the up- and down-type Higgs doublets, as in the 2HDM that will be a part of our low-energy model. The $\mathbf{126}$ decomposes as
\begin{eqnarray}
 \mathbf{126}&=&\Delta_L\left(10,3,1\right)_{422}\oplus \Delta_R\left(\overline{10},1,3\right)_{422}\nn\\ && \oplus \Sigma\left(15,2,2\right)_{422} \oplus\;\Xi^\prime \left(6,1,1\right)_{422}\;. 
\end{eqnarray}

The breaking sequence of the gauge symmetry of the model is schematically given as
%
%\begin{widetext}
\begin{equation}
\label{eq:chain}
 %\quad 
 \mathrm{SO}(10) \,\underset{\left<\mathbf{ 54}_H\right>}{\xrightarrow{M_U}}\,G_{422D} \,\underset{\left<\mathbf{126}_H\right>}{\xrightarrow{M_\mathrm{PS}}}\, G_{321}\,\mbox{(SM)} \,\underset{\left<\mathbf{10}_H\right>}{\xrightarrow{M_Z}}\, G_{31}\;,
\end{equation}
%\end{widetext}
%
%\\
where we have used the notation
\begin{eqnarray}
G_{422D}&\equiv& \mathrm{SU}(4)_{\mathcal{C}}\otimes \mathrm{SU}(2)_L\otimes \mathrm{SU}(2)_R\otimes  D\;,\cr
G_{321}&\equiv& \mathrm{SU}(3)_{\mathcal{C}}\otimes \mathrm{SU}(2)_L\otimes \mathrm{U}(1)_{Y}\;,\cr
G_{31}&\equiv& \mathrm{SU}(3)_{\mathcal{C}}\otimes \mathrm{U}(1)_{Q}\;.
\end{eqnarray}
Here $D$ denotes the $D$-parity, or left-right symmetry, a $\mathbb{Z}_2$ symmetry that maintains the complete equivalence of the left and right sectors~\cite{Chang:1983fu,Maiezza:2010ic}. The first stage of the spontaneous symmetry-breaking occurs through the Pati-Salam singlet in $\mathbf{54}_H$ acquiring a VEV at the unification scale $M_U$. This singlet is even under $D$ parity and, therefore, the resulting symmetry group is $G_{422D}$.\footnote{A $\mathbf{210}_{H}$, instead of the $\mathbf{54}_{H}$, can also be used to break the $\mathrm{SO}(10)$ symmetry, in which case the $D$-parity will also be broken at $M_U$, since the Pati-Salam singlet in $\mathbf{210}_{H}$ is odd under it. We however choose to use the $\mathbf{54}_{H}$ in order for our model to align with the one in Ref~\cite{Babu:2015bna}.} 

In the second step, the breaking of the $G_{422D}$ into the SM gauge group $G_{321}$ is realised when the SM singlet, contained in $\Delta_R(\overline{10},1,3)_{422}$ of $\mathbf{126}_H$, acquires a VEV at the energy scale $M_{\mathrm{PS}}$, which also yields a Majorana mass for the right-handed neutrino. In the last stage of the symmetry-breaking,  the SM doublets contained in $\Phi(1,2,2)_{422}$ of the complex $\mathbf{10}_H$ acquire a VEV at the EW scale. Additionally, the SM doublets in the $\Sigma(15,2,2)_{422}$ of the $\mathbf{126}_H$ acquire VEVs as well, with their contributions correcting the otherwise problematic fermion mass and mixing relations, to yield a realistic model. Since these contributions are controlled by the factor $M_{\mathrm{PS}}/M_{\Sigma}$, the mass of $\Sigma(15,2,2)_{422}$ ($M_{\Sigma}$) should be around the Pati-Salam breaking scale, $M_{\mathrm{PS}} \equiv \left\langle \mathbf{126}_{H}\right\rangle$, for them to have the desired magnitudes~\cite{Bajc:2005zf,Babu:2015bna}. The fermion mass matrices for the up- and down-type quarks, charged leptons, Dirac neutrinos, and Majorana neutrinos are given as
\begin{eqnarray}
\label{massmatrices}
M_U &=& v^{u}_{10} Y_{10}+v^{u}_{126} Y_{126}\;,\quad M_D= v^{d}_{10}  Y_{10}+v^{d}_{126} Y_{126}\nonumber\\
& &\quad \quad M_E= v^{d}_{10} Y_{10}-3v^{d}_{126} Y_{126}\;,\nonumber\\
M_{\nu_D}&=& v^{u}_{10} Y_{10}-3v^{u}_{126} Y_{126}\;, \quad M_{\nu_{M}}= \sigma Y_{126}\;,
\end{eqnarray}
where $v^{u}_{10}, v^{d}_{10}$ are the VEVs of the two complex doublets in the (complex) $ \mathbf{10}_H$, $v^{u}_{126}$ and $v^{d}_{126}$ are VEVs of the two complex doublets in $\Sigma\left(15,2,2\right)_{422}$ of the $\mathbf{126}_H$, while $\sigma$ denotes the VEV of the SM singlet in $\mathbf{126}_H$.

The potential term, $\propto \mathbf{126}_H \cdot \overline{\mathbf{126}}_H \cdot \mathbf{126}_H \cdot \mathbf{10}_H$ responsible for the above fixing, as well as the potential terms corresponding to the singlet field $S$, which breaks the U(1)$_{\mathrm{PQ}}$ symmetry, introduce mixing in the mass matrices. Therefore, the doublets $(1,2,\frac{1}{2})_{321}$ and triplets $(3,1,-\frac{1}{3})_{321}$ in $\mathbf{10}_H$ and $\mathbf{126}_H$ mix among their own kinds. A fine-tuning of the parameters appearing in the mixing matrices is commonly adopted to ensure that, out of this combination, the masses of only the SM doublets lie at the EW scale, while all the other doublets and triplets, which can induce proton-decay-mediating interactions, remain at the intermediate and the GUT-breaking scales. As noted earlier, we deviate from the last assumption in this paper.

The mass-separation of the components of a large multiplet is generally known as the splitting problem in the literature. The mass terms of the doublets and triplets in a given $\mathbf{10}_H$ are identified with the same parameters, up to $\mathcal{O}(1)$. The relevant potential terms are given as~\cite{Babu:2015bna}
\begin{eqnarray}
\label{potential}
\mathcal{V}&=&m^2|\mathbf{10}_{H}|^2+\eta^\prime \mathbf{10}_{H}^*\cdot \mathbf{54}_{H}\cdot \mathbf{10}_{H}+\frac{\eta_0} {2} \mathbf{54}_{H}^2 |\mathbf{10}_{H}|^2 \nonumber\\
&+&\eta_2 \mathbf{10}_{H}^*\cdot \mathbf{54}_{H}\cdot \mathbf{54}_{H}\cdot \mathbf{10}_{H}+\chi|\mathbf{10}_{H}|^2 |S|^2\nonumber\\
&+&\frac{\gamma_1}{4!} \mathbf{10}_{H}^*\cdot \mathbf{\overline{126}}_{H}\cdot \mathbf{126}_{H}\cdot \mathbf{10}_{H}\nonumber\\
&+&\frac{\gamma_2}{4!} \mathbf{10}_{H}^*\cdot \mathbf{126}_{H}\cdot \mathbf{\overline{126}}_{H}\cdot \mathbf{10}_{H}\nonumber\\
&+&\eta_1 \mathbf{126}_H \cdot \overline{\mathbf{126}}_H \cdot \mathbf{126}_H \cdot \mathbf{10}_H...
\end{eqnarray}
$\mathbf{54}_{H}$ acquires a VEV
\begin{eqnarray}
\left<\mathbf{54}_{H}\right>=\text {diag}(-2 / 5\; \mathbf{I}_{6\times6}, \;3 / 5\; \mathbf{I}_{4\times4})\omega\;,
\end{eqnarray}
 breaking the $\mathrm{SO}(10)$ symmetry, where $\omega\sim M_U$. The VEVs of the singlets in $\mathbf{54}_{H}$ and $\mathbf{126}_{H}$ are given as $\left<\mathbf{1}_{54}\right>=-\sqrt{12/5}\; \omega$ and  $\left<\mathbf{1}_{126}\right>=\sigma/\sqrt{2}$ ($\simeq M_{\mathrm{PS}}$), and the VEV of $S$, which itself is a singlet, becomes  $\left<S\right>=v_s/\sqrt{2}\simeq M_{\mathrm{PQ}}\simeq f_a$. The diagonal terms in the mass-squared matrices of the doublets and triplets of $\mathbf{10}_{H}$ are given as~\cite{Babu:2015bna}
\begin{eqnarray}
\begin{aligned}
M^2_{\xi_1 } & =-\frac{2}{5} \eta^\prime \omega+\frac{6}{5} \eta_0 \omega^2+\frac{4}{25} \eta_2 \omega^2+\gamma_1 \sigma^2+\frac{1}{2} \chi v_s^2+m^2, \\
M^2_{\xi_2 }  & =-\frac{2}{5} \eta^\prime \omega+\frac{6}{5} \eta_0 \omega^2+\frac{4}{25} \eta_2 \omega^2+\gamma_2 \sigma^2+\frac{1}{2} \chi v_s^2+m^2,\\
M^2_{\phi_1}  & =\frac{3}{5} \eta^\prime \omega+\frac{6}{5} \eta_0 \omega^2+\frac{9}{25} \eta_2 \omega^2+\gamma_1 \sigma^2+\frac{1}{2} \chi v_s^2+m^2,\\
M^2_{\phi_2} & =\frac{3}{5} \eta^\prime \omega+\frac{6}{5} \eta_0 \omega^2+\frac{9}{25} \eta_2 \omega^2+\gamma_2 \sigma^2+\frac{1}{2} \chi v_s^2+m^2.\nonumber
\end{aligned}
\hspace*{-0.5cm}\\
\end{eqnarray}

The similarity of these terms makes it difficult to split the masses of the members of the multiplet, keeping a doublet light and others at the GUT scale. The mixing with other doublets and triplets, introduced by the $\eta_1$ term in Eq.~(\ref{potential}) could ameliorate the splitting problem through fine-tuning. Instead, we adopt the picture that whatever mechanism keeps the Higgs doublet light brings other members of the $\mathbf{10}_{H}$ down to the TeV scale even after the mixing. Note that we still have the seemingly fine-tuned situation but assume that $\mathbf{10}_{H}$ remains light altogether. We deal with the issue of proton stability through a discrete symmetry to forbid the diquark couplings of our light leptoquarks, as will be mentioned in the next section. 
 
To sum up, we assume that the choice of parameters that keeps one of the doublets at the EW scale also brings the masses of the other components of the complex $\mathbf{10}_H$, \textit{i.e.}, the other Higgs doublet and the two leptoquarks, down to around the same scale. From a minimalist point of view, it is therefore conceivable that the elements necessary to resolve all the anomalies persisting in the experimental data come from a multiplet that resides altogether at the TeV scale. The picture presented in this paper hence originates from an absence of large splitting of the doublet-triplet mass parameters within a multiplet. It, therefore, constitutes a good motivation to keep the entire complex $\mathbf{10}_H$ light, providing a rationale for the existence of new physics around the TeV scale window in the SO(10) framework. 

%The fermion mass relations of the $\mathrm{SO}(10)$ model with the same GUT Yukawa sector as ours (but with only the SM at the TeV-scale) were studied in Ref.~\cite{Babu:2015bna}. Since the overall effects of a small change in the particle content at the TeV-scale on the GUT-scale fermion mass relations are expected to be small~\cite{Aydemir:2019ynb}, we have no reason to suspect any significant deviation from that analysis.
% of Ref.~\cite{Babu:2016bmy}, and leave this examination for future work.
 
\subsection{The Lagrangian at low energy}
%: 2HDM with leptoquarks of $\mathbf{\textit{S}_1}$ type}

The fields with the same quantum numbers as the two Higgs doublets and the two $S_1$-type leptoquarks, given in Eq.~(\ref{10H-decomp}), are denoted in the low-energy Lagrangian below as $H_1$ and $H_2$, and $\widetilde{L}_1$ and $\widetilde{L}_2$, respectively.\footnote{We will re-denote the physical leptoquark states after mass-mixing at the TeV scale by $Q_1$ and $Q_2$ in Sec. \ref{sec:pheno}.} Thus, at the TeV scale, we effectively have a 2HDM augmented by two leptoquarks. For this model to be phenomenologically consistent though, we have to address the problem of the possible decay of the proton via our leptoquarks. The proton stability, in general, could be ensured through various symmetry mechanisms such as the utilisation of a U(1) symmetry~\cite{Cox:2016epl,Pati:1974yy} or a discrete symmetry~\cite{Bauer:2015knc}. The relevant operators could also be suppressed by mechanisms such as the one discussed in Ref.~\cite{Dvali:1995hp}, or they could be completely forbidden for geometrical reasons~\cite{Aydemir:2018cbb}. 

In our framework, proton stability is ensured by a $\mathbb{Z}_2$ symmetry, which is commonly introduced in an ad-hoc manner (see for instance Refs.~\cite{Bauer:2015knc,Aydemir:2019ynb}). However, the U(1)$_{\mathrm{PQ}}$ symmetry in our scenario offers a more compelling, and arguably less artificial, alternative. The required discrete symmetry might be a leftover one from the breaking of the U(1)$_{\mathrm{PQ}}$, and has previously been identified as $(-1)^{3B}$~\cite{Dasgupta:2013cwa}, with $B$ being the baryon number. We impose that this symmetry is valid in our framework as well. The corresponding charges are then given as $(q,\, \widetilde{L}_1,\,\widetilde{L}_2)\rightarrow (- q,-\widetilde{L}_1,\,-\widetilde{L}_2)$. The rest of the SM particle content is not charged under this symmetry. Therefore, the diquark couplings of the leptoquarks are forbidden and the proton stability is not a concern in the perturbative framework. 
%(A small violation of this symmetry will arise due to the non-perturbative effects, as in the SM.)  

In a 2HDM, the coupling of both the Higgs fields to the SM fermions can lead to flavour-changing neutral currents. To avoid these, the most general approach taken is to enforce a $\mathbb{Z}_2^H$ symmetry on the Lagrangian, so that each type of fermions only couples to one of the doublets \cite{Glashow:1976nt,Paschos:1976ay}. This leads to `Types' I-IV of the 2HDM. In our scenario, the 2HDM (augmented by two leptoquarks) that automatically arises from the $\mathrm{SO}(10)$ Lagrangian given in Eq.~(\ref{eq:UVLagrangian}), is of the Type II, wherein the $H_2$ couples only to the $u$-type quarks, while $H_1$ couples to the $d$-type quarks and the charged leptons. In other words, the $\mathbb{Z}_2^H$ symmetry principle corresponding to the Type-II 2HDM is naturally satisfied. Therefore, the Lagrangian relevant for our phenomenological analysis is given by

\begin{eqnarray}
\label{eq:lagrangianLQ}
\mathcal{L}_{S}^{Y}&\supset&  
\lt(\boldsymbol{\Lambda}^{1L}_{q\ell}\bar{Q}^c i\tau_2 L+\boldsymbol{\Lambda}^{1R}_{q\ell}\bar{u}_R^c e_R\rt)\widetilde{L}_1^{*} \nonumber\\
&+&\lt(\boldsymbol{\Lambda}^{2L}_{q\ell}\bar{Q}^c i\tau_2 L+\boldsymbol{\Lambda}^{2R}_{q\ell}\bar{u}_R^c e_R\rt)\widetilde{L}_2 +\mathrm{H.c.}\;,\\
\mathcal{L}^{Y}_H&\supset& \bar{Q}\left( Y_1^d H_1\right)d_R
+\bar{Q}(Y_2^u  \widetilde {H}_2)u_R\nonumber\\
&+& \bar{L}\left( Y_1^e H_1\right)e_R
+ \bar{L} (Y_2^{\nu} \widetilde {H}_2)\nu_R + \mathrm{H.c.}\;,
\end{eqnarray}
where $Q$ and $L$ are the SM quark and lepton doublets (of a given family), respectively, $\boldsymbol{\Lambda}^{iL/iR}_{q\ell}$ are the leptoquark coupling matrices in flavour space, and $\psi^c=C\bar{\psi}^T$ are charge-conjugate spinors. Note also that the inclusion of the $\widetilde{L}_{1,2}$ can affect the stability of the EW vacuum via loop effects~\cite{Bandyopadhyay:2016oif}.
% The relevant discussion is found in Ref.

The leptoquark-Higgs coupling terms in the potential are given as
\begin{eqnarray}
\label{eq:PotLQH}
-\mathcal{L}_{SH}^P&\supset&
\lambda_{11} |\widetilde{L}_1|^2 |H_1|^2+\lambda_{12} |\widetilde{L}_1|^2 |H_2|^2\nonumber\\
& +&\lambda_{21} |\widetilde{L}_2|^2 |H_1|^2+\lambda_{22} |\widetilde{L}_2|^2 |H_2|^2\;,
\end{eqnarray}
while the rest of the (CP-conserving) potential reads
\begin{eqnarray}
\label{eq:PotRest}
-\mathcal{L}^P&\supset& \bar{M}_{1}^{2} |\widetilde{L}_1|^2 + \bar{M}_{2}^2 |\widetilde{L}_2|^2
-[\bar{M}_{12}^2 \widetilde{L}_1^\dagger \widetilde{L}_2 +\,\text{H.c.} ]\nonumber\\
&+& M_1^2 |H_1|^2+ M_2^2 |H_2|^2 
-[M_{12}^2 H_1^\dagger H_2 +\,\text{H.c.} ]\nonumber\\
&+& \frac{1}{2}\bar{\lambda}_1 (\widetilde{L}_1^\dagger \widetilde{L}_1)^2 + \frac{1}{2}\bar{\lambda}_2 (\widetilde{L}_2^\dagger \widetilde{L}_2)^2\nonumber\\
&+& \bar{\lambda}_3 (\widetilde{L}_1^\dagger \widetilde{L}_1)(\widetilde{L}_2^\dagger \widetilde{L}_2)
+ \bar{\lambda}_4 (\widetilde{L}_1^\dagger \widetilde{L}_2)(\widetilde{L}_2^\dagger \widetilde{L}_1)\nonumber\\
&+& \left[\frac{1}{2}\bar{\lambda}_5 (\widetilde{L}_1^\dagger \widetilde{L}_2)^2 + \,\text{H.c.} \right]\nonumber\\
&+& \frac{1}{2}\lambda_1 (H_1^\dagger H_1)^2 + \frac{1}{2}\lambda_2 (H_2^\dagger H_2)^2\nonumber\\
&+& \lambda_3 (H_1^\dagger H_1)(H_2^\dagger \widetilde{L}_2)
+ \lambda_4 (H_2^\dagger H_2)(H_1^\dagger \widetilde{L}_1)\nonumber\\
&+& \left[\frac{1}{2}\lambda_5 (H_1^\dagger H_2)^2 
+ \,\text{H.c.} \right]\;.
\end{eqnarray}

Evidently, the Lagrangian given in Eq.~(\ref{eq:lagrangianLQ}) should be understood in the effective field theory context. The Higgs and leptoquark couplings in this TeV-scale Lagrangian are induced from the original $\mathrm{SO}(10)$ Yukawa couplings. Each of these Yukawa couplings is generated by a linear combination of the unification-scale operators, and gets modified due to the mixing effects induced by the scalar fields' Yukawa couplings to the three chiral families of $\mathbf{16}_F$. It is indeed this rich structure that enables the realisation of a fermion mass spectrum consistent with that of the SM expected at the unification scale~\cite{Babu:2015bna}. The modification to the renormalisation group (RG) running of the SM Yukawa couplings due to the inclusion of $\boldsymbol{\lambda}_{L/R}$ does not lead to substantial changes in this mass spectrum, as discussed in the next section, and hence the main message of ~\cite{Babu:2015bna} is valid in our case as well.

%%%%%%%%%%%%%%%%%%%%%%%%%%%%%%%%%%%%%%%%%%%%%%%%%%%%%%%%%%%%
\subsection{Gauge coupling unification \label{sec:unification}}

The new particle content at the TeV scale does not lead to the unification of the SM gauge couplings directly, as can be easily verified from the usual formalism given below. We, therefore, adopt the scheme of gauge-coupling unification with a single intermediate step of breaking the Pati-Salam symmetry into the SM group at the scale $M_{\mathrm{PS}}$, as detailed earlier.

For the purpose of our analysis, the one-loop RG running is sufficient. The formalism of this running for the gauge symmetry-breaking sequence shown in Eq.~(\ref{eq:chain}) is given in the Appendix \ref{sec:RG-running}. The scalar content and the RG coefficients in the corresponding energy intervals are given in Table \ref{a1B}, where we label the energy intervals in between the breaking scales
$[M_Z,M_{\mathrm{PS}}]$ and $[M_{\mathrm{PS}},M_U]$ as
\begin{eqnarray}
\mathrm{I}  & \;:\; & [M_Z,\;M_{\mathrm{PS}}]\;,\quad G_{321}  \;(\mathrm{SM})\;,\cr
\mathrm{II} & \;:\; & [M_{\mathrm{PS}},\;M_U]\;,\quad  G_{422D} \;.
\label{IntervalNumber}
\end{eqnarray}

\begin{table*}
\begin{center}
\begin{tabular}{c|c|c}
\hline
$\vphantom{\Big|}$ Interval & Scalar content for model & RG coefficients
\\
\hline\hline
$\vphantom{\Biggl|}$ II
&  $\Phi(1,2,2)$, $\Xi(6,1,1)$,\vspace{-0.05cm}&\vspace{-0.2cm}
\\
& $\Sigma(15,2,2)$, $\Delta_R(\overline{10},1,3)$, &$\left[ a_{4},a_{L},a_{R}\right]=\left[1,\dfrac{26}{3},\dfrac{26}{3}\right]$\\
&$\ \Delta_L(10,3,1)$
& $\vphantom{\Big|}$ 
\\
\hline
$\vphantom{\Biggl|}$   I
& $\phi_1\left(1,2,\frac{1}{2}\right)_{321}$, $\xi_1 \left(3,1,-\frac{1}{3}\right)_{321}$
&\quad\;\; \;\;$\left[a_{3},a_{2},a_{1}\right]=\left[\dfrac{-20}{3},-3,\dfrac{65}{9}\right]$ \\
&\vspace{-0.4cm}&\\
& $\phi_2\left(1,2,-\frac{1}{2}\right)_{321}$, $\xi_2\left(\bar{3},1,\frac{1}{3}\right)_{321}$ \vspace{0.1cm}& 
\\
\hline
\end{tabular}
\label{a1B}
\caption{The scalar content and the RG coefficients in the energy intervals I and II.}
\end{center}
\end{table*}

\begin{figure}[t!]
\centering
\captionsetup[subfigure]{labelformat=empty}
\includegraphics[width=\columnwidth]{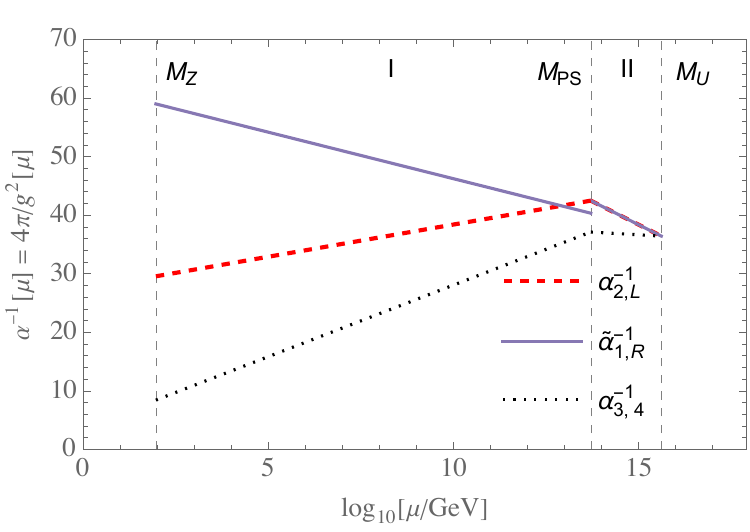}%\hspace{0.4cm}
\caption{Running of the gauge couplings. Note that $\widetilde{\alpha}^{-1}_1\equiv \frac{3}{5}\alpha^{-1}_1$. The discontinuity at $M_{\mathrm{PS}}$ is due to the boundary conditions given in Eq.~\eqref{Matching}.}
\label{RGrunning}
\end{figure}

The central values of the low-energy parameters, used as the boundary conditions in the RG running 
(in the $\overline{\mathrm{MS}}$ scheme) \cite{ParticleDataGroup:2024cfk}, are
$\alpha^{-1} = 127.95$, $\alpha_s = 0.118$, and $\sin^2\theta_W = 0.2312$ at $M_Z=91.2\,\mathrm{GeV}$, which translate to $g_1 = 0.357$, $g_2 = 0.652$, and $g_3 = 1.219$. 
As usual, the coupling constants are all required to remain in the perturbative regime during the evolution from $M_Z$ to $M_U$. Thus, once the RG coefficients in each interval are specified,
the scales $M_U$ and $M_{\mathrm{PS}}$, along with the value of $\alpha_U$, are uniquely determined through Eqs.~(\ref{unification-relations}) and (\ref{A6}). This yields
\begin{eqnarray}
\label{UV-predictions}
&&\log_{10}\left(\frac{M_U}{\mathrm{GeV}}\right)=15.6\;,\quad \log_{10}\left(\frac{M_{\mathrm{PS}}}{\mathrm{GeV}}\right)=13.7\;,\nonumber\\
&&\qquad\qquad\qquad\qquad \alpha_U^{-1}=36.5\;,
\end{eqnarray}
and the unification of the couplings is displayed in Fig.~\ref{RGrunning}.\footnote{The energy scales found in Eq.~(\ref{UV-predictions}) are the same as the ones in the model studied in Ref.~\cite{Aydemir:2019ynb}. This is because, although the RG coefficient $a_i$ are different in each energy interval for the two models, their combinations entering in the Eq.~(\ref{unification-relations}), which determines the energy scales, turn out to be the same, which is not the case for $\alpha_U^{-1}$ and the corresponding Eq.~(\ref{A6}).}

%%%%%%%%%%%%%%%%%%%%%%%%%%%%%%%%%%%%%%%%%%%%%%%%%%%%%%%%%%%%%%%%%%%%%%%

As mentioned previously, the proton-decay-mediating couplings of the $\widetilde{L}_1$ are $\widetilde{L}_2$ are forbidden by an assumed discrete symmetry. However, we do not make any assumptions regarding the other potentially dangerous relevant operators. Thus, it is necessary to inspect whether the predictions given in Eq.~(\ref{UV-predictions}) are compatible with the current experimental bounds. The most recent and stringent bound on the lifetime of the proton is $\tau_p^{\rm obs} >1.6\times 10^{34}$ years~\cite{Miura:2016krn}. In our model, the proton-decay can also potentially be mediated by the super-heavy gauge bosons residing in the $\mathbf{45}$ adjoint representation of the $\mathrm{SO}(10)$. For the corresponding operators, $\tau_p\sim M_U^4/m_p^5 \alpha_U^2$~\cite{Langacker:1980js}, where $m_p$ is the proton mass. We obtain $\tau_p\sim 10^{34}$ years, which is consistent with the observation up to $O(1)$. Additionally, the scale $M_{\mathrm{PS}}$ determines the expected mass values for the proton-decay-mediating color triplets. A naive analysis~\cite{Altarelli:2013aqa} shows that the current bounds on the $\tau_p$ require $M_{\mathrm{PS}}\gtrsim 10^{11}$~GeV, again consistent with the results given in Eq.~(\ref{UV-predictions}).

%%%%%%%%%%%%%%%%%%%%%%%%%%%%%%%%%%%%%%%%%%%%%%%%%%%%%%%%%%%%%%%%%%%%%%%

\section{Low-energy parameters and observables}
\label{sec:pheno}

From Eqs.~\eqref{eq:lagrangianLQ}-\eqref{eq:PotRest}, we collect the complete set of free parameters of the model as
\begin{align}
&\big\{\Lambda^{1L}_{q\ell},\,\Lambda^{1R}_{q\ell},\,\Lambda^{2L}_{q\ell},\,\Lambda^{2R}_{q\ell},\, Y_1^d,\,Y_2^u,\,Y_1^e,\,Y_2^\nu,\nonumber\\
&\quad \lambda_{11},\,\lambda_{12},\,\lambda_{21},\,\lambda_{22},\,\bar{M}_1^2,\,\bar{M}_2^2,\,\bar{M}_{12}^2,\\
&\quad M_1^2,\,M_2^2,\,M_{12}^2,\,\bar{\lambda}_{1\cdots 5},\,\lambda_{1\cdots 5}\big\}\;\nonumber.
\end{align}

After EW symmetry-breaking, the scalar doublets $H_1$ and $H_2$ are defined in terms of their respective VEVs $v_1$ and $v_2$, the physical Higgs states $h$, $H$, $A$ and $H^\pm$, and the Goldstone bosons $G$ and $G^\pm$, as
\begin{equation}
H_1=\frac{1}{\sqrt{2}}\left(\begin{array}{c}
\displaystyle \sqrt{2}\left(G^+ c_\beta -H^+ s_\beta\right)  \\
\displaystyle v_1-h s_\alpha+H c_\alpha+i\left( G c_\beta-A s_\beta \right)
\end{array}
\right)\;,
\end{equation}
\begin{equation}
H_2=\frac{1}{\sqrt{2}}\left(\begin{array}{c}
\displaystyle \sqrt{2}\left(G^+ s_\beta +H^+c_\beta\right)  \\
\displaystyle v_2+h c_\alpha+Hs_\alpha+i\left( G s_\beta+A c_\beta \right)
\end{array}
\right)\;,
\end{equation}
where $c_x$ and $s_x$ stand for $\cos(x)$ and $\sin(x)$, respectively, with $\beta$ ($\equiv \tan^{-1}[v_1/v_2]$) and $\alpha$ being the angles rotating the CP-odd and CP-even interaction states, respectively, into physical Higgs states. 

As a result, the bare masses $M_1$ and $M_2$ can be replaced by $v_1$ and $v_2$ using the tadpole conditions for the Higgs potential. In addition, the Higgs quartic couplings $\lambda_{1\cdots 5}$ can be traded for the physical masses of the Higgs bosons $(m_h,\,m_H,\,m_A,~{\rm and}~m_{H^\pm})$ and the mixing parameter $s_{\beta-\alpha} \equiv\sin(\beta-\alpha)$ as inputs. The Yukawa couplings of the four Higgs bosons are then calculated in the Type-II 2HDM in terms of the mixing angles as
\begin{gather*}
Y_h^u=g_u\frac{c_\alpha}{s_\beta}\;,~~Y_h^{d/\ell}=-g_{d/\ell}\frac{s_\alpha}{c_\beta}\;,\\
Y_H^u=g_u\frac{s_\alpha}{s_\beta}\;,~~Y_H^{d/\ell}=g_{d/\ell}\frac{c_\alpha}{c_\beta}\;,\\
Y_A^u=\frac{g_u}{t_\beta}\;,~~Y_A^{d/\ell}=g_{d/\ell}t_\beta\;,
\end{gather*}
where $t_\beta\equiv \tan\beta$, and $g_f=m_f/v$, with $v=\sqrt{v_1^2+v_2^2}$. 

At least one of the Higgs bosons in our model ought to have its mass and Yukawa couplings consistent with the one observed at the LHC. We identify it with the lighter of the two neutral scalars, by fixing $m_h= 125$\,GeV and $s_{\beta-\alpha}=1.0$. The latter corresponds to the so-called alignment limit~\cite{Bernon:2015wef} of the Type-II 2HDM, wherein the $h$ has exactly SM-like couplings to the gauge bosons, since $g_h^V\propto s_{\beta-\alpha}$, while $g_H^V\propto c_{\beta-\alpha}$. In order that the Higgs sector is consistent with the direct search exclusion limits from the LHC, we fixed $M_{12}^2 = (150\,{\rm GeV})^2$,\, $t_\beta=12$, and $m_H=m_A=m_{H^\pm}= 1$\,TeV. To ensure this consistency, we nevertheless tested this Higgs sector parameter with the {\tt HiggsBounds-v5.10.0}~\cite{Bechtle:2020pkv}) program. Equal masses of all the heavy Higgs states also imply that the oblique parameters $S$,\,$T$, and $U$ are always $\sim 0$, and hence satisfy the 95\% confidence level (CL) limits from the 2024 PDG report~\cite{ParticleDataGroup:2024cfk}. 

As for the leptoquark sector, since this analysis is focused specifically on its contributions to the anomalous flavour observables rather than its own collider phenomenology, we assume the minimal low-energy scenario. We set $\bar{M}_{12}^2=\bar{\lambda}_1=\bar{\lambda}_2=\bar{\lambda}_3=\bar{\lambda}_4=\bar{\lambda}_5=0$, so that $\bar{M}^2_1=m^2_{Q_1}$ and $\bar{M}^2_2=m^2_{Q_2}$, where $Q_1$ ($Q_2$) is the lighter (heavier) of the two leptoquark mass eigenstates. This also allows us to retain the leptoquark couplings $\Lambda^{1L}_{q\ell},\,\Lambda^{1R}_{q\ell},\,\Lambda^{1L}_{q\ell}$, and $\Lambda^{2R}_{q\ell}$ as the input free-parameters at the EW scale, but we switch their respective notations to $\lambda^{Q_1L}_{q\ell},\,\lambda^{Q_1R}_{q\ell},\,\lambda^{Q_2L}_{q\ell}$, and $\lambda^{Q_2R}_{q\ell}$ from here onward.  

The contribution of the $S_1$-type $\widetilde{L}$ to $a_\mu$ can be approximated by \cite{2005.04352}
\begin{equation}
\Delta a_\mu\simeq -\frac{N_c}{8\pi^2}\frac{m_t m_\mu}{m_{\widetilde{L}}^2}V_{tb}\lambda^{\widetilde{L}L}_{32}\lambda^{\widetilde{L}R}_{32}\left[\frac{7}{6}+\frac{2}{3}\log x_t\right]\,,
\end{equation}
%(see, e.g.,~\cite{Marzocca:2021azj}. 
where $m_t$ ($m_\mu$) is the top (muon) mass, $x_t = m_t^2/m_{\widetilde{L}}^2$, and $V_{tb}$ is the relevant CKM matrix element. Unlike $a_\mu$, the observables 
\ba
R_D &\simeq & R_D^{\rm SM}\Big(|1 + C_V^L|^2 + 1.09|C_S^L|^2 + 0.75|C_T^L|^2 \nonumber\\
&+& 1.54{\mathrm Re}[(1+C_V^L)(C_S^L)^*] \nonumber\\
&+& 1.04{\mathrm Re}[(1+C_V^L)(C_T^L)^*]
+ 1.5{\mathrm Re}(C_{RL}^{\tau 3}+C_{LL}^{\tau 3}) \nonumber\\
&+& 1.0|C_{RL}^{\tau 3}+C_{LL}^{\tau 3}|^2 \Big)\;,
\ea
and
\ba
R_{D^*} &\simeq & R_{D^*}^{\rm SM}\Big(|1 
+ C_V^L|^2 + 0.05|C_S^L|^2 + 16.27|C_T^L|^2 \nonumber \\
&-& 0.13{\mathrm Re}[(1+C_V^L)(C_S^L)^*]\nonumber \\
&-& 5.0{\mathrm Re}[(1+C_V^L)(C_T^L)^*]
+ 0.12{\mathrm Re}(C_{RL}^{\tau 3}+C_{LL}^{\tau 3}) \nonumber\\ 
&+& 0.05|C_{RL}^{\tau 3}+C_{LL}^{\tau 3}|^2 \Big)\;,
\ea
receive contributions from the $\widetilde{L}$ as well as from the $H^\pm$ at the tree level. The Wilson coefficients corresponding to the $\widetilde{L}$ are given as~\cite{2005.04352}
\ba
C_V^L &=& \frac{1}{2\sqrt{2} G_F  m_{\widetilde{L}}^2}\frac{\lambda_{23}^{\widetilde{L}L}\lambda_{33}^{\widetilde{L}L}}{2V_{cb}}\;,\nonumber\\
C_S^L &=& -\frac{1}{2\sqrt{2} G_F m_{\widetilde{L}}^2}\frac{\lambda_{33}^{\widetilde{L}L}\lambda_{23}^{\widetilde{L}R}}{2V_{cb}}\;,~~
C_T^L = - \frac{1}{4} C_S^L\;,
\ea  
while those corresponding to the $H^\pm$ are~\cite{Chen:2021vzk} 
\be
C_{RL}^{\tau 3}\approx\frac{m_b m_\tau}{m_{H^\pm}^2}\left(\frac{2}{t_\beta}^2 + 1\right)\;,~~C_{LL}^{\tau 3}=\frac{m_c m_\tau}{m_{H^\pm}^2}
\ee
(neglecting the highly subdominant $C_{LR}^{\tau 3}$ and $C_{RR}^{\tau 3}$ terms).

\begin{figure*}[tbp]
\captionsetup[subfigure]{labelformat=empty}
\subfloat[\quad\quad(a)]{\includegraphics[width=0.49\linewidth]{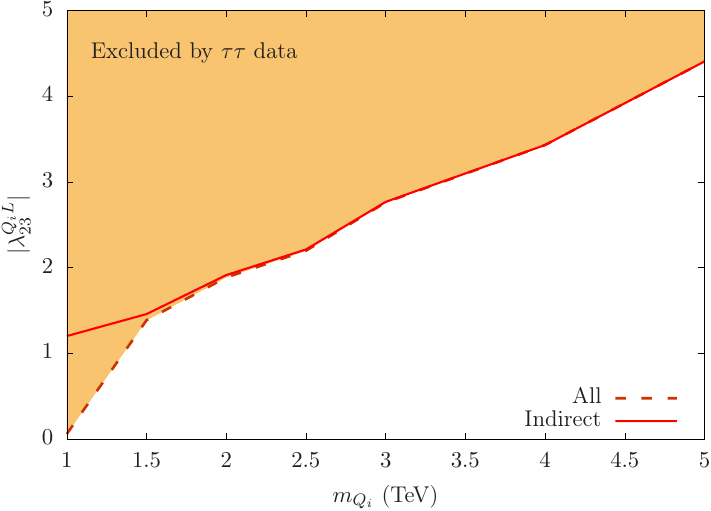}\label{fig:lm23l}}\hspace{0.2cm}
\subfloat[\quad\quad(b)]{\includegraphics[width=0.49\linewidth]{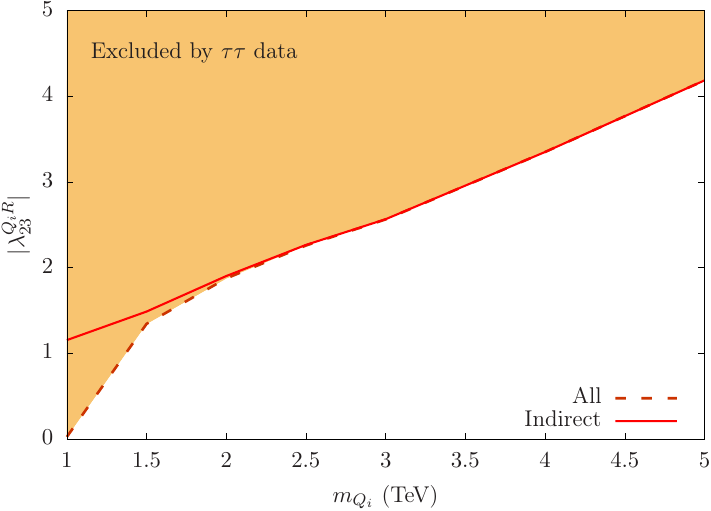}\label{fig:lm23r}}
\caption{Limits on $Q_i$ couplings from ATLAS di-$\tau$ data~\cite{Aad:2020zxo} obtained following Refs.~\cite{Mandal:2018kau, Bhaskar:2021pml}. The solid lines show the limits obtained by ignoring the contributions from the single- and pair-productions of the $Q_i$ to the $\tau\tau$ data (the regions above the lines are excluded). The dashed lines show how the limits change if we do not neglect these processes but ignore the presence of all other couplings except the one under consideration.}
\label{fig:lhclims}
\end{figure*}

In our model, both the $Q_1$ and $Q_2$ should enter these Wilson coefficients, and their couplings relevant for this study thus include $\big\{\lambda^{Q_1L}_{23},\,\lambda^{Q_1L}_{32},\,\lambda^{Q_1L}_{33},\,\lambda^{Q_1R}_{23},\,\lambda^{Q_1R}_{32}\big\}$ and $\big\{\lambda^{Q_2L}_{23},\,\lambda^{Q_2L}_{32},\,\lambda^{Q_2L}_{33},\,\lambda^{Q_2R}_{23},\,\lambda^{Q_2R}_{32}\big\}$. Besides $a_\mu$ and $R_{D^{(*)}}$, combinations of these couplings are additionally constrained by the experimental results for
\begin{equation} 
\hspace*{-0.2cm} R_{K^{(*)}} \equiv \dfrac{\mathrm{BR}(B\rightarrow K^{(*)}\mu^+\mu^-)}{\mathrm{BR}(B\rightarrow K^{(*)}e^+e^-)}\propto \lambda^{\widetilde{L}L*}_{23}\lambda^{\widetilde{L}L}_{33}|\lambda^{\widetilde{L}L}_{32}|^2\;,
\end{equation} 
and 
\begin{equation} 
R_K^{\nu} \equiv \frac{{\rm BR}(B\to K \nu \nu)}{{\rm BR}(B\to K \nu \nu)_{\rm SM}} \propto \lambda_{23}^{\widetilde{L}L} \lambda_{33}^{\widetilde{L}L}\;.
\end{equation}
Furthermore, certain other observables, such as BR$(\tau\to\mu\gamma)$, BR$(\tau\to 3\mu)$, and BR$(B_c\to\tau\nu)$ also have a (slightly more involved) dependence on these couplings (see, e.g., \cite{2005.04352} for their complete expressions). 

\section{Numerical analysis of the 2HDM+2$\widetilde{L}$}
\label{sec:numeric}

In order to verify that the values of $\Delta a_\mu$ and $R_{D^{(*)}}$ predicted by our model can be consistent with their experimental measurements given in Eqs. (2) and (3), respectively, we analyzed some representative low-energy configurations of its parameter space. We required these configurations to pass the following three criteria to be selected as benchmark points (BPs) for further analysis.

\subsection{Yukawa RG running and perturbativity}\label{sec:Yukawa}

All of our ten $\lambda^{Q_{1,2}}$ couplings, in addition to the $t$-quark Yukawa coupling, do not hit a Landau pole at high energies. For a parameter space point to qualify as a BP, these couplings should thus be small enough to stay in the perturbative realm up to the GUT scale. We ignore the rest of the SM Yukawa runnings, as they are considerably smaller. The 1-loop RG equations for the couplings of our concern are given in Appendix \ref{sec:Yukawa}. Also, as commonly done in literature, we run the RG equations from the EW scale all the way to the GUT scale, $M_U$, ignoring the effects of the intermediate symmetry-breaking scale, $M_{\rm PS}$. These effects are also expected to be insignificant~\cite{Meloni:2014rga} since the running is logarithmic, and $M_{\rm PS}$ is quite close to the $M_U$ in our model. 

\subsection{Direct search limits from the LHC}
\label{subsec:lhcbounds}

As discussed briefly in the previous section, the non-zero $\lambda^{\widetilde{L}X}_{23}$, $\lambda^{\widetilde{L}X}_{32}$, and $\lambda^{\widetilde{L}L}_{33}$ couplings, with $X=L,\,R$, of each of the $\widetilde{L}_{1,2}$ in our model would lead to some interesting signatures at the LHC, involving decay modes such as
\begin{equation*}
    \widetilde{L}\to c\tau,\;s\nu,\;t\mu,\;b\nu,\;{\rm and}\;t\tau,
\end{equation*}
where we have suppressed the neutrino flavours, as the LHC cannot mutually distinguish between them. Depending on the magnitudes of the couplings, the BRs in one or more of these modes can be sizeable. As a result, the pair-production of a given $\widetilde{L}$ at the LHC would give rise to different possibilities (third-generation quarks and leptons, third-generation quarks and second-generation leptons, etc.). Similarly, it would also have sizeable single-production, leading to more possible signatures. Some of these channels have been probed (see, e.g., Refs.~\cite{Sirunyan:2018kzh, Aaboud:2019bye, Aad:2020iuy, CMS:2020wzx}), while the prospects of some others at the high-luminosity LHC have also been studied~\cite{Chandak:2019iwj,Bhaskar:2020gkk,Bhaskar:2021gsy,Bhaskar:2023xkm,Ali:2023kss}. The current direct-search exclusion bounds on a scalar leptoquark reach up to $1.7$\,TeV for maximal BR in some modes (see, e.g., Table 1 of Ref.~\cite{Bhaskar:2021pml}).

However, even such heavy leptoquarks can leave indirect signatures at the LHC, especially if they have large Yukawa couplings. A pair of quarks can interchange a $\widetilde{L}$ in the $t$-channel and produce a lepton pair. As a result, with a large $\lambda^{\widetilde{L}L}_{23}$ coupling, $\widetilde{L}$ could be responsible for a significant number of di-tau events at the LHC through the $c\bar c\to \tau^+\tau^-$ process. The process is highly sensitive to the unknown coupling, as the $\widetilde{L}$-exchange diagram contributes to the cross section proportionally to the fourth power of $\lambda^{\widetilde{L}L}_{23}$. Moreover, the $\tilde{L}$-mediated process and the SM $c\bar c\to Z^{*}/\gamma^{*} \to \tau^+\tau^-$ process undergo destructive interference, $\propto (\lambda^{\widetilde{L}L}_{23})^2$. Overall, for $\lambda^{\widetilde{L}L}_{23}\sim\mathcal O(1)$, the process would noticeably affect the $\tau\tau$ distributions. Hence, the current dilepton data can constrain the parameter space of a model~\cite{Mandal:2018kau, Bhaskar:2021pml}, as does the monolepton + missing transverse energy (MET) data for the $\widetilde{L}$-mediated $qq^\prime\to \ell\nu$ process. However, since Ref.~\cite{Mandal:2018kau} has shown the limits from the dilepton data to be stronger, we only consider these here.
Moreover, the $\tau\tau jj$ direct-search bounds from the LHC~\cite{ATLAS:2023kek,CMS:2023bdh}, recasted following the method described in Ref.~\cite{Bhaskar:2023ftn}, are similarly much weaker than the indirect bounds shown in Fig.~\ref{fig:lhclims}.

Of the five non-zero couplings of each $\widetilde{L}$ in our analysis, the three $\lambda^{\widetilde{L}X}_{3i}$ couplings are unaffected by the dilepton data. This is because a charge\,-$1/3$ $\widetilde{L}$ can couple to a charged lepton only with a $t$-quark through these couplings. Hence, one needs $t\bar t$ pairs in the initial state to produce $\mu^+\mu^-/\tau^+\tau^-$ pairs through $\widetilde{L}$-exchange. But since the parton density function (PDF)of the $t$-quark is negligible, these couplings would not be constrained by the $\mu^+\mu^-$ or $\tau^+\tau^-$ (or even the $\mu/\tau$+MET) data. Finally, according to Ref.~\cite{Bhaskar:2023ftn}, search results for $\widetilde{L}$ pair-production in the $\ell t\ell t$ channels \cite{ATLAS:2024huc,CMS:2022nty} are also relatively much less constraining.

We show the exclusion limits on the absolute magnitudes of $\lambda^{\widetilde{L}L}_{23}$ and $\lambda^{\widetilde{L}R}_{23}$ from the ATLAS $\tau\tau$ data~\cite{Aad:2020zxo} in Figs.~\ref{fig:lm23l} and \ref{fig:lm23r}, respectively. (Recall that these two couplings contribute to the $R_{D^{(*)}}$ observables, but not to $\Delta a_\mu$.) These limits are obtained by considering two degenerate LQs having $\lm_{23}^{Q_1L}=\lm_{23}^{Q_2L}$ (as is the case for our BPs 1, 3, and 5, described in the next section). Even though, we include the contributions from the single- and pair-production processes to the limits for completeness, these are relatively minor for the mass range we consider. For BPs 2 and 4, for which $\lm_{23}^{Q_1L}\neq \lm_{23}^{Q_2L}$, the coupling values are well within the allowed range. We have computed all the necessary cross-sections using the {\tt Universal FeynRules Output} (UFO)~\cite{Degrande:2011ua} model files from Ref.~\cite{Mandal:2018kau} in {\tt MadGraph5}~\cite{Alwall:2014hca} with the NNPDF23LO~\cite{Ball:2012cx} PDF set. We incorporated the QCD $k$-factor to compute the pair-production cross sections~\cite{Mandal:2015lca}, and combined various production events following Refs.~\cite{Mandal:2012rx,Mandal:2015vfa,Mandal:2016csb}. 

\subsection{Indirect constraints}
\label{subsec:flavour}

As noted in the previous section, a $\widetilde{L}$ can contribute to certain other flavor-changing decays, and its properties (i.e., mass and couplings) are therefore strongly constrained by the results of the experimental searches for these processes. A parameter space configuration was therefore acceptable only if it satisfied the constraints listed in Table~\ref{tab:const}.

\begin{table}
\begin{center}
{\begin{tabular}{|c|c|}
\hline
Observable & Range/limit \\ \hline
\hline
$R_K$ & $0.91 \,\text{-}\, 1.09$ \\ 
$R_K^\nu$ & $ < 2.7$~\cite{Belle:2017oht} \\
BR$(\tau\to\mu\gamma)$ & $< 4.4\times 10^{-8}$ \\
BR$(\tau\to 3\mu)$ &  $< 2.1\times 10^{-8}$ \\
BR$(B_c\to\tau\nu)$ & $< 0.1$ \\\hline
BR$(B\to X_s \gamma)$ & \{3.17\,-\,3.47\}\,$\times 10^{-4}$~\cite{HFLAV:2016hnz} \\
BR$(B_s \to \mu\mu)$ & \{2.15\,-\,3.85\}\,$\times 10^{-9}$ ~\cite{Aaij:2017vad} \\
BR$(B_u\to \tau\nu)$ & \{0.87 \,-\,1.25\}\,$\times 10^{-4}$~\cite{HFLAV:2016hnz} \\ 
\hline
\end{tabular}}
\end{center}
\caption{\label{tab:const} Experimental constraints imposed on various observables in our analysis. For measurements, the $1\sigma$ statistical and systematic uncertainties, when given separately in the referenced result, were added in quadrature, while the upper limits quoted are at the 95\% CL.}
\end{table}

\section{Resolution of flavor anomalies}
\label{sec:results}

In order to calculate the predictions of the various observables in the 2HDM+2$\widetilde{L}$, we incorporated it in the {\tt Mathematica} package {\tt SARAH-v4.14.4}~\cite{Staub:2008uz,Staub:2013tta,Goodsell:2014bna,Goodsell:2015ira,Staub:2015kfa,Goodsell:2016udb,Braathen:2017izn}. This package automatically calculates the expressions for all the interaction vertices in order to write down model files for the {\tt FORTRAN} code {\tt SPheno-v4.0.5}~\cite{Porod:2003um,Porod:2011nf} for carrying out phenomenological studies. For a given input parameter space configuration, {\tt SPheno} computes the mass spectra of all the particles in the model, as well as the decay BRs for some of them. It additionally computes the model predictions for most of the observables under consideration here, with the exception of $R_{D^{(*)}}$, $R_K^{\nu}$, and BR$(B_c\to\tau\nu)$, which were estimated by implementing their expressions from~\cite{2005.04352} in a local numerical code. 

The parameter values of the five BPs we analyse here are listed in Table~\ref{tab:BPs}. Given that, despite fixing many of the model's parameters, the number of possible configurations of the 12 free ones meeting our requirements can potentially be very large (an exhaustive collection of which is beyond the scope of this study), we have identified these BPs using the following criteria.

\begin{itemize}
\item They satisfy all the constraints described in the previous section. 
\item The values of $\Delta a_\mu$ and $R_{D^{(*)}}$ lie almost exactly at the bottom of their respective $\pm$1$\sigma$ ranges, given in Eqs.~(2) and (3), i.e., $R_{D^*}\simeq 0.275$, and $\Delta a_\mu\simeq 20.1\times 10^{-10}$.\footnote{The magnitudes of the couplings needed for $R_{D^*}$ to reach the lower edge of its current experimental 1$\sigma$ range are overall larger than for $R_{D}$. The former is therefore more constraining than the latter, and it is typically not possible to obtain the minimum allowed of values of both of these for a unique set of the relevant couplings.} This means that the magnitude of any of the couplings for a given BP can not be reduced any further without adjusting some other coupling to obtain the desired values of these observables.  
\item $m_{Q_1}=m_{Q_2}$, simply to reduce the number of free parameters to adjust. We will, however, briefly discuss the implications of a sizeable splitting between the masses of $Q_1$ and $Q_2$ later. Also, $m_{Q_{1,2}}\geq 2$\,TeV ensures consistency with the strongest direct search bound noted in the previous section, irrespective of their mass-degeneracy and the sizes of their couplings.
\item For BPs 1, 3, and 5, the corresponding couplings of $Q_1$ and $Q_2$ have exactly the same value. Furthermore, $\lambda^{Q_iL}_{23}=-\lambda^{Q_iR}_{23}$ (since consistency with the $R_{D^{(*)}}$ measurements warrants one of the relevant couplings to be negative), and $\lambda^{Q_iL}_{32}=\lambda^{Q_iR}_{32}$. These BPs thus represent the simplest possible scenario in the model, wherein the number of free parameters needed to meet all the enforced requirements is minimum. (We emphasize that none of these assumptions have been made on theoretical grounds.)
\item BP2 is representative of an alternative possible scenario, in which $Q_1$ has couplings large enough to yield the minimum allowed values of $\Delta a_\mu$ and $R_{D^*}$, while the contribution of $Q_2$ is negligible. On the other hand, given the larger $m_{Q_{1,2}}$ for BP4, $Q_2$ contributes sizeably to $R_{D^*}$, since the couplings of $Q_1$ cannot get sufficiently large without violating perturbativity at the GUT scale. For both these points, again only for simplification, we set $\lambda^{Q_iL}_{23}=-\lambda^{Q_iR}_{23}$ and $\lambda^{Q_2L}_{32}=\lambda^{Q_2R}_{32}$.
\end{itemize}

The purpose of these BPs is to analyze the modifications needed in the various couplings for them to be consistent with the enforced constraints as the mass(es) of $Q_1$ or (and) $Q_2$ is (are) increased. The numerical values of each of the couplings given in the table are dictated by their perturbativity at the GUT scale, as well as the interplay between the experimental bounds on different observables. Thus, while $\lambda^{Q_{1,2}L}_{23}$ ($=-\lambda^{Q_{1,2}R}_{23}$) can all have substantially large magnitudes to satisfy $R_{D^{(*)}}$ for BPs 1 and 3 and still stay perturbative, $\lambda^{Q_{1,2}L}_{33}$ are restricted to much smaller values by the $R_K^\nu<2.7$ limit. Similarly, $\lambda^{Q_1L}_{32}$ can have values as large as 0.57 and 0.6, respectively, for BPs 2 and 4, but $\lambda^{Q_1R}_{32}$ is strongly constrained by the upper limit on BR$(\tau\to\mu\gamma)$, as a result of which a slightly enhanced contribution is also needed from $Q_2$ to achieve $\Delta a_\mu \geq 20.1\times 10^{-10}$ for these two BPs. 

\begin{table}[tbp]
\begin{center}
\begin{tabular}{|l|c|c|c|c|c|}
%\begin{tabular*}{\columnwidth}{|l @{\extracolsep{\fill}} |c|c|c|c|}
\hline
        & BP1 & BP2 & BP3 & BP4 & BP5 \\ \hline
$m_{Q_{1,2}}$ (TeV) & 2.0 & 2.0 & 2.5 & 2.5 & 3.0 \\
\hline
\hline
$\lambda^{Q_1L}_{23}$ & \multirow{2}{*}{0.53}    & 0.71   & \multirow{2}{*}{0.66}   & 0.78   & \multirow{2}{*}{0.8}   \\
$\lambda^{Q_2L}_{23}$ &                          & 0.01   &                        & 0.42   &                         \\ %\hdashline
\hline
$\lambda^{Q_1L}_{33}$ & \multirow{2}{*}{0.032}    & 0.1   & \multirow{2}{*}{0.038} & 0.12   & \multirow{2}{*}{0.046}  \\
$\lambda^{Q_2L}_{33}$ &                          & 0.01   &                        & 0.015   &        \\
\hline
$\lambda^{Q_1R}_{23}$ & \multirow{2}{*}{--0.53}  & --0.71 & \multirow{2}{*}{--0.66} & --0.78 & \multirow{2}{*}{--0.8} \\ 
$\lambda^{Q_2R}_{23}$ &                          & --0.01 &                        & --0.42 &                         \\ 
\hline
\hline
$\lambda^{Q_1L}_{32}$ & \multirow{2}{*}{0.078}   & 0.57   & \multirow{2}{*}{0.092} & 0.6   & \multirow{2}{*}{0.105}  \\
$\lambda^{Q_2L}_{32}$ &                          & 0.01  &                        & 0.02  &                         \\
\hline
$\lambda^{Q_1R}_{32}$ & \multirow{2}{*}{0.078}   & 0.02   & \multirow{2}{*}{0.092} & 0.026   & \multirow{2}{*}{0.105}  \\ 
$\lambda^{Q_2R}_{32}$ &                          & 0.01  &                        & 0.02  &                         \\  
\hline
\hline
$\Gamma_{Q_1}$ (GeV)    & \multirow{2}{*}{34.4}   & 86.8    & \multirow{2}{*}{66.5}   & 128   & \multirow{2}{*}{117}    \\
$\Gamma_{Q_2}$ (GeV)    &                        & 0.032   &                        & 26.4    &                         \\ 
\hline
\hline
BR($Q_1\to s\nu_\tau $) & \multirow{2}{*}{0.326} & 0.232  & \multirow{2}{*}{0.326} & 0.237  & \multirow{2}{*}{0.327}  \\
BR($Q_2\to s\nu_\tau $) &                        & 0.126  &                        & 0.332  &                         \\
\hline
BR($Q_1\to c\tau $)     & \multirow{2}{*}{0.636} & 0.453  & \multirow{2}{*}{0.637} & 0.464   & \multirow{2}{*}{0.638}  \\
BR($Q_2\to c\tau $)     &                        & 0.255  &                        & 0.648  &                         \\
\hline
BR($Q_1\to b\nu_\mu $)  & \multirow{2}{*}{0.007}  & 0.149   & \multirow{2}{*}{0.006} & 0.14  & \multirow{2}{*}{0.006}  \\ 
BR($Q_2\to b\nu_\mu $)  &                        & 0.126  &                        & 0.00  &                         \\ 
\hline
BR($Q_1\to t\mu $)      & \multirow{2}{*}{0.014} & 0.147  & \multirow{2}{*}{0.012} & 0.139  & \multirow{2}{*}{0.011}  \\
BR($Q_2\to t\mu $)      &                        & 0.247  &                        & 0.00  &                         \\
\hline
%\hline
%$m_{Q_2}$ (TeV) & 2.0 & 2.0 & 2.4 & 2.4 & 2.8 \\
\end{tabular}
\caption{\label{tab:BPs} Values of the input parameters, and of some $Q_1$ and $Q_2$ decay observables, for each of the five selected BPs.}
\end{center}
\end{table}

\begin{figure*}[tbp]
\captionsetup[subfigure]{labelformat=empty}
\subfloat[\hspace{0.5cm}(a)]{\includegraphics[width=0.43\linewidth]{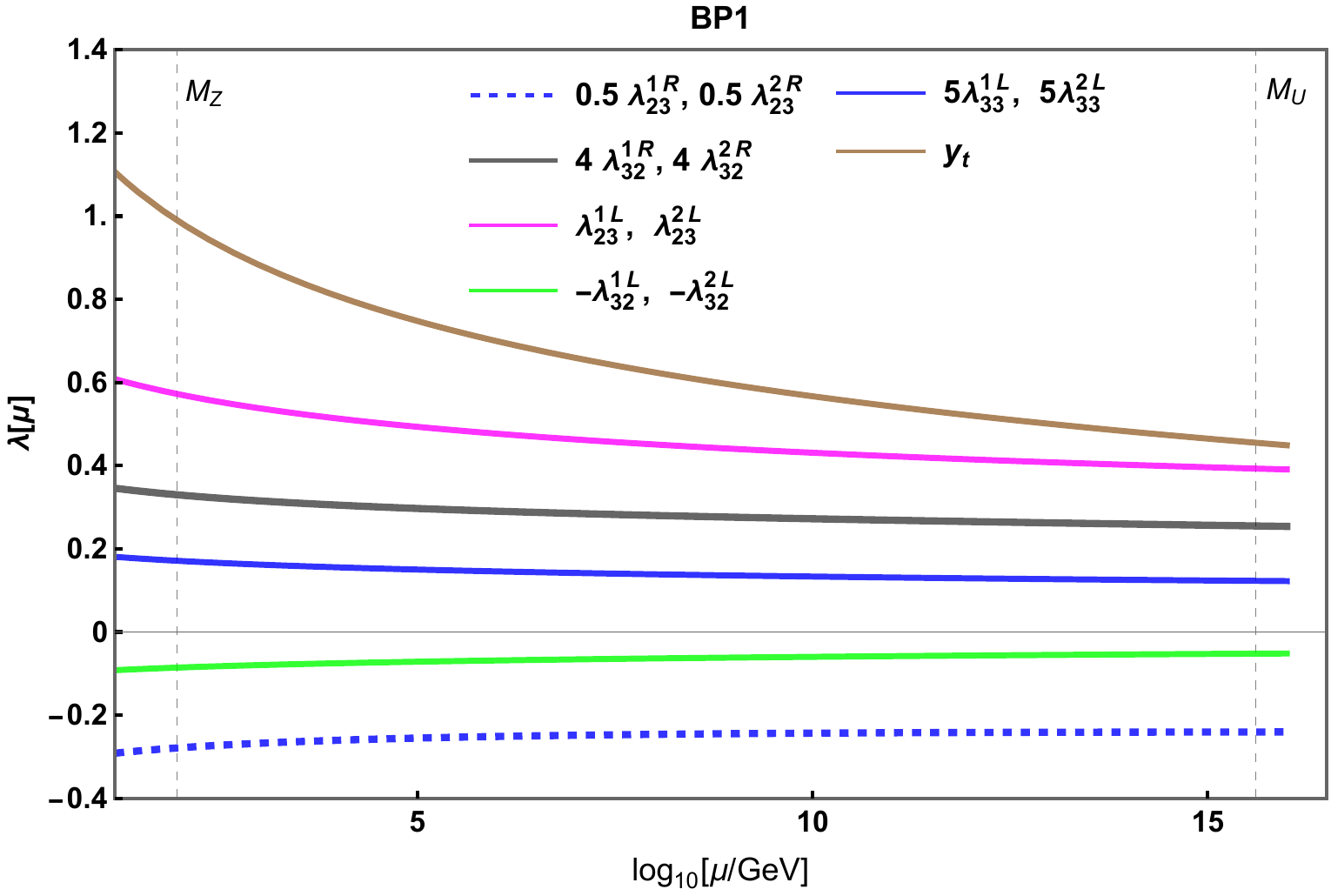}\label{fig:YukawaBP1}}\hspace{2.2cm}
\subfloat[\quad\quad(d)]{\includegraphics[width=0.43\linewidth]{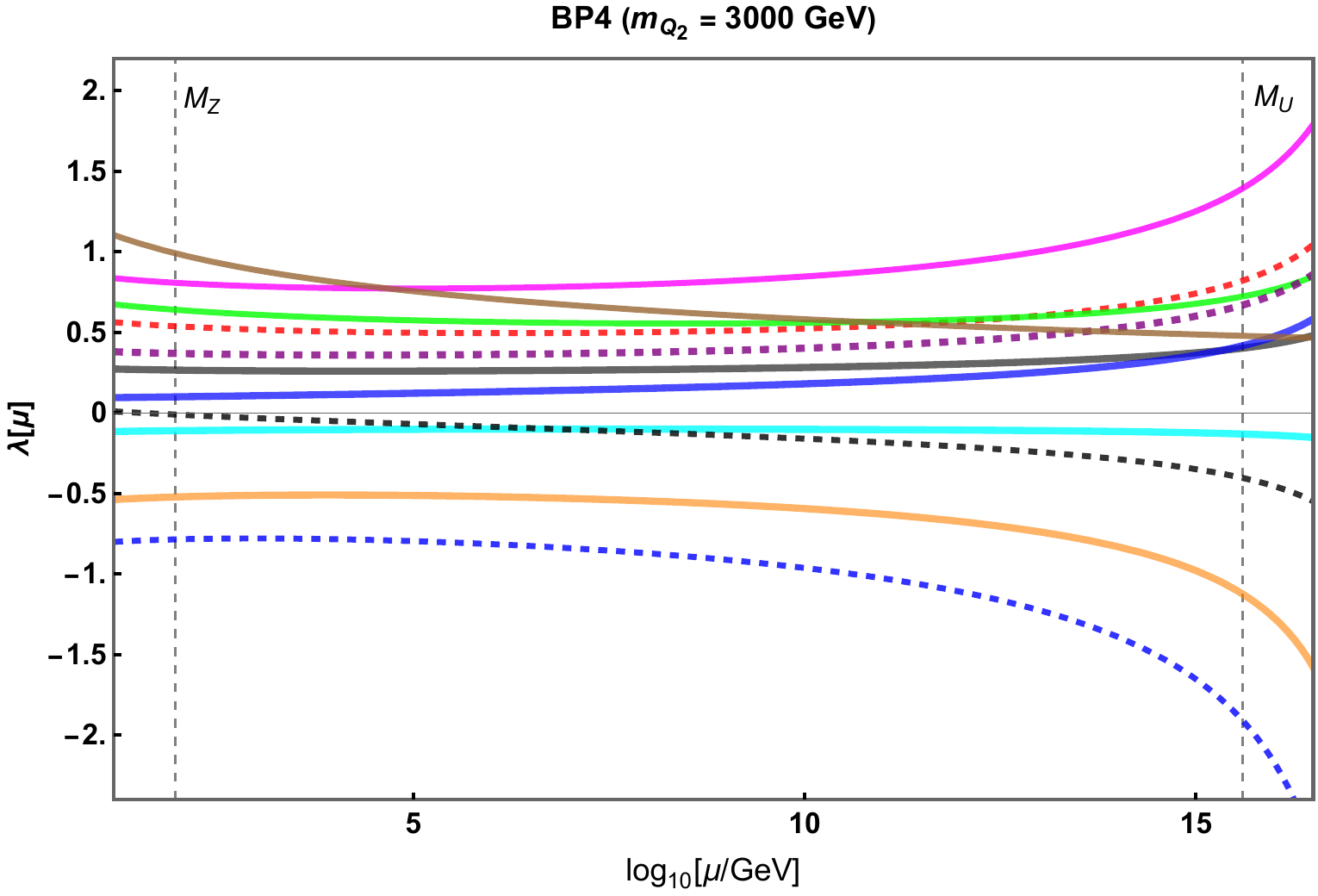}\label{fig:YukawaBP42}}\\
\subfloat[\hspace{-2.2cm}(b)]{\includegraphics[width=0.58\linewidth]{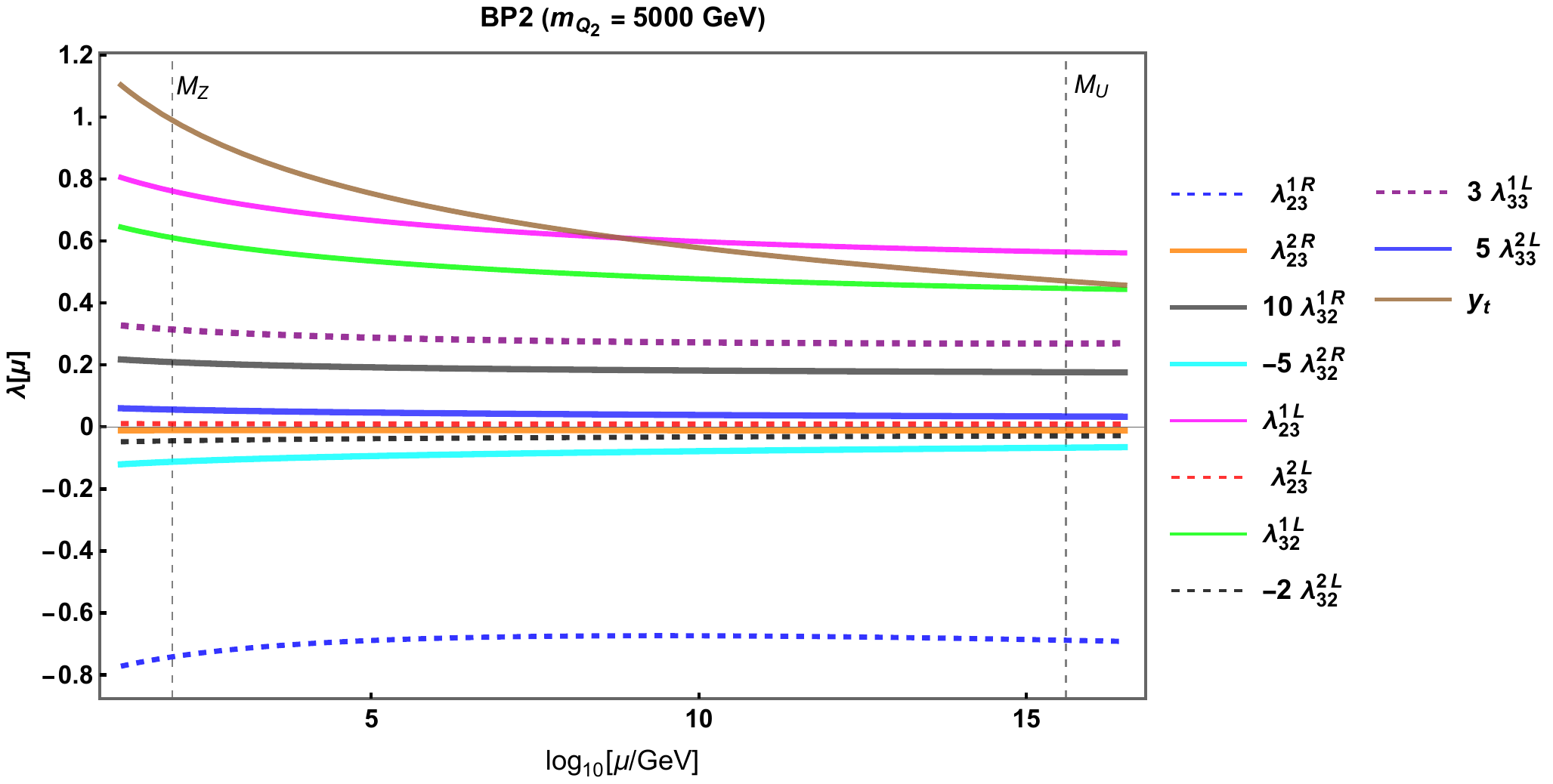}\label{fig:YukawaBP2}}\hspace{-0.4cm}
\subfloat[\quad\quad(e)]{\includegraphics[width=0.42\linewidth]{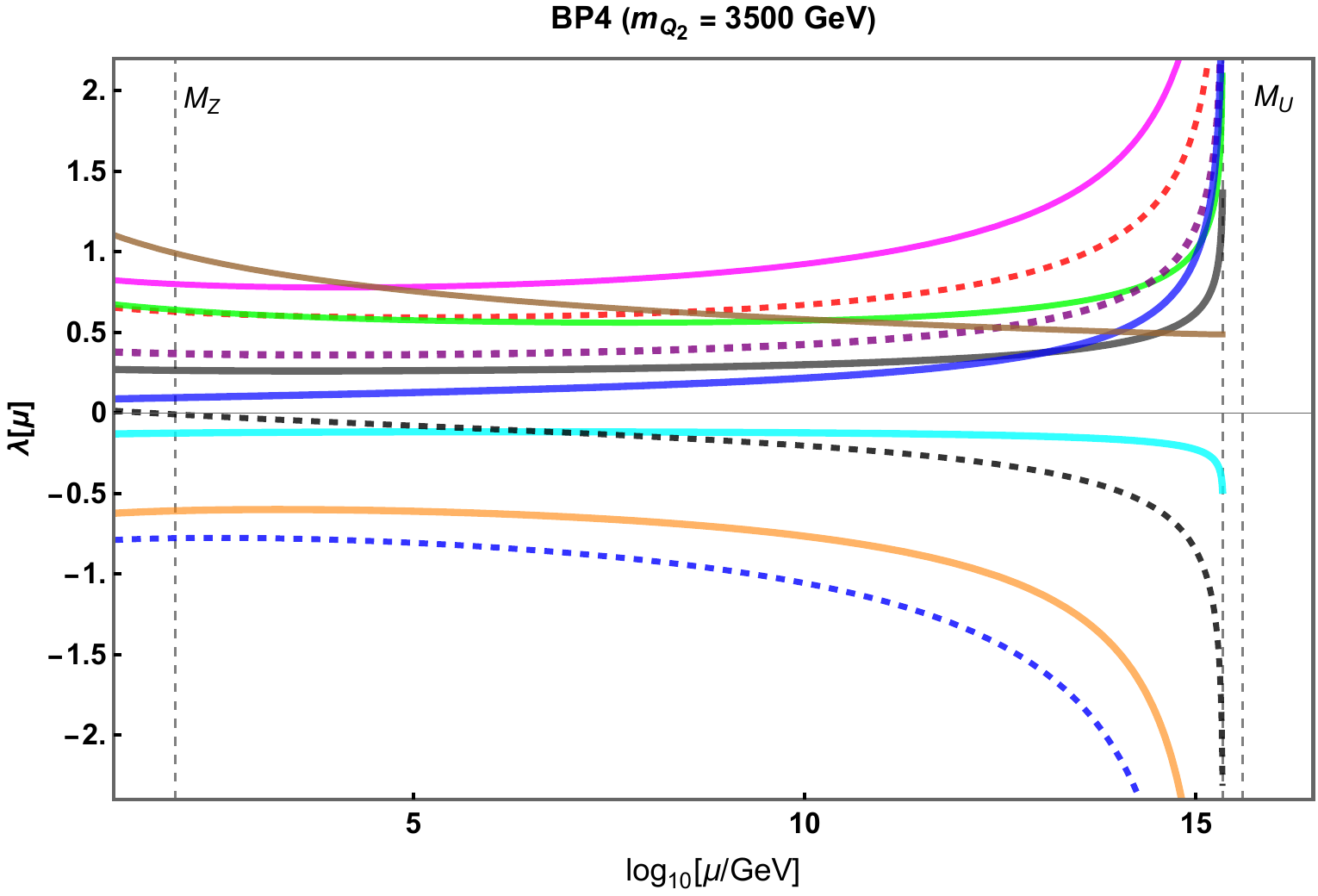}\label{fig:YukawaBP43}}\\
\subfloat[\quad\quad(c)]{\includegraphics[width=0.43\linewidth]{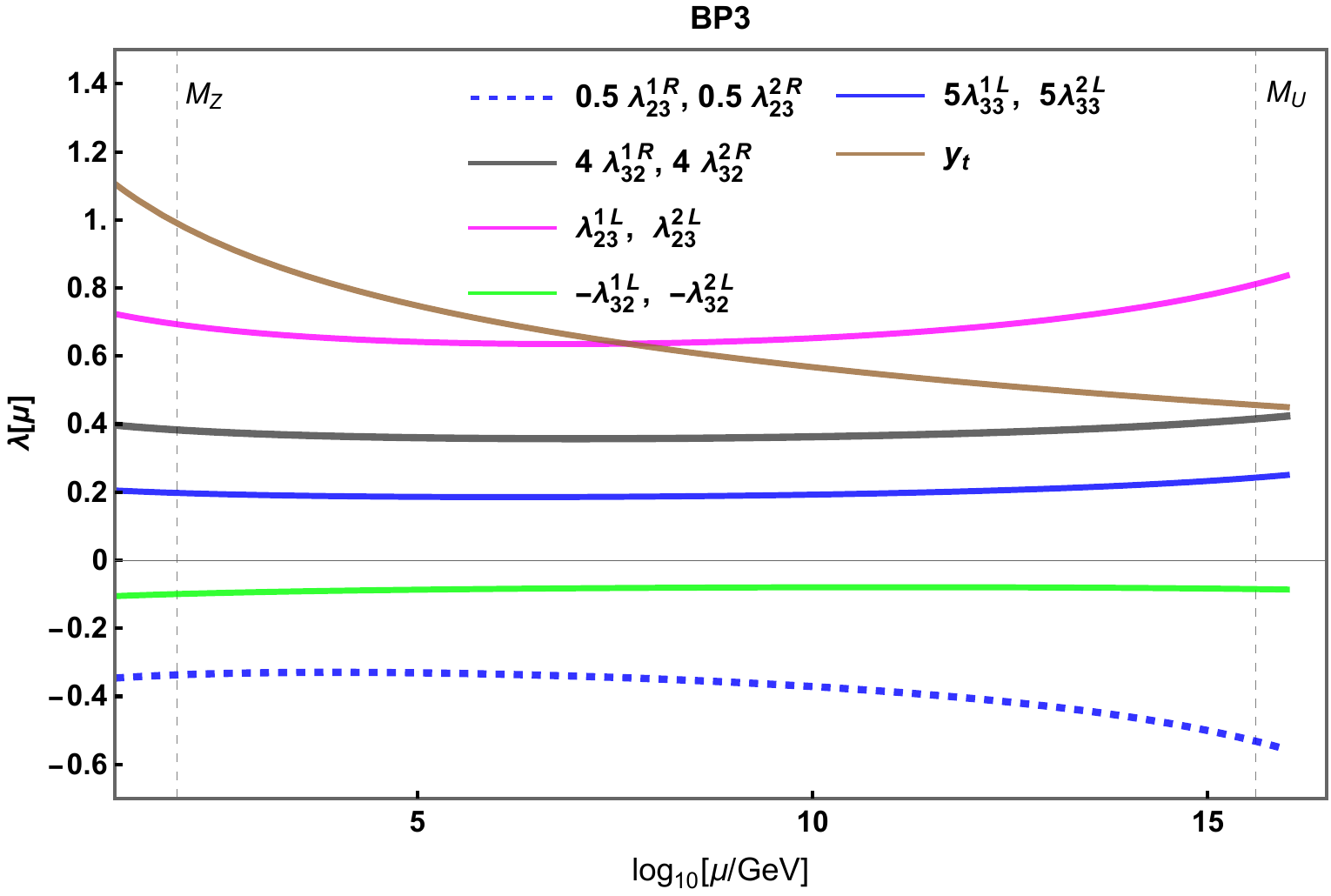}\label{fig:YukawaBP3}}\hspace{2.2cm}
\subfloat[\quad\quad(f)]{\includegraphics[width=0.42\linewidth]{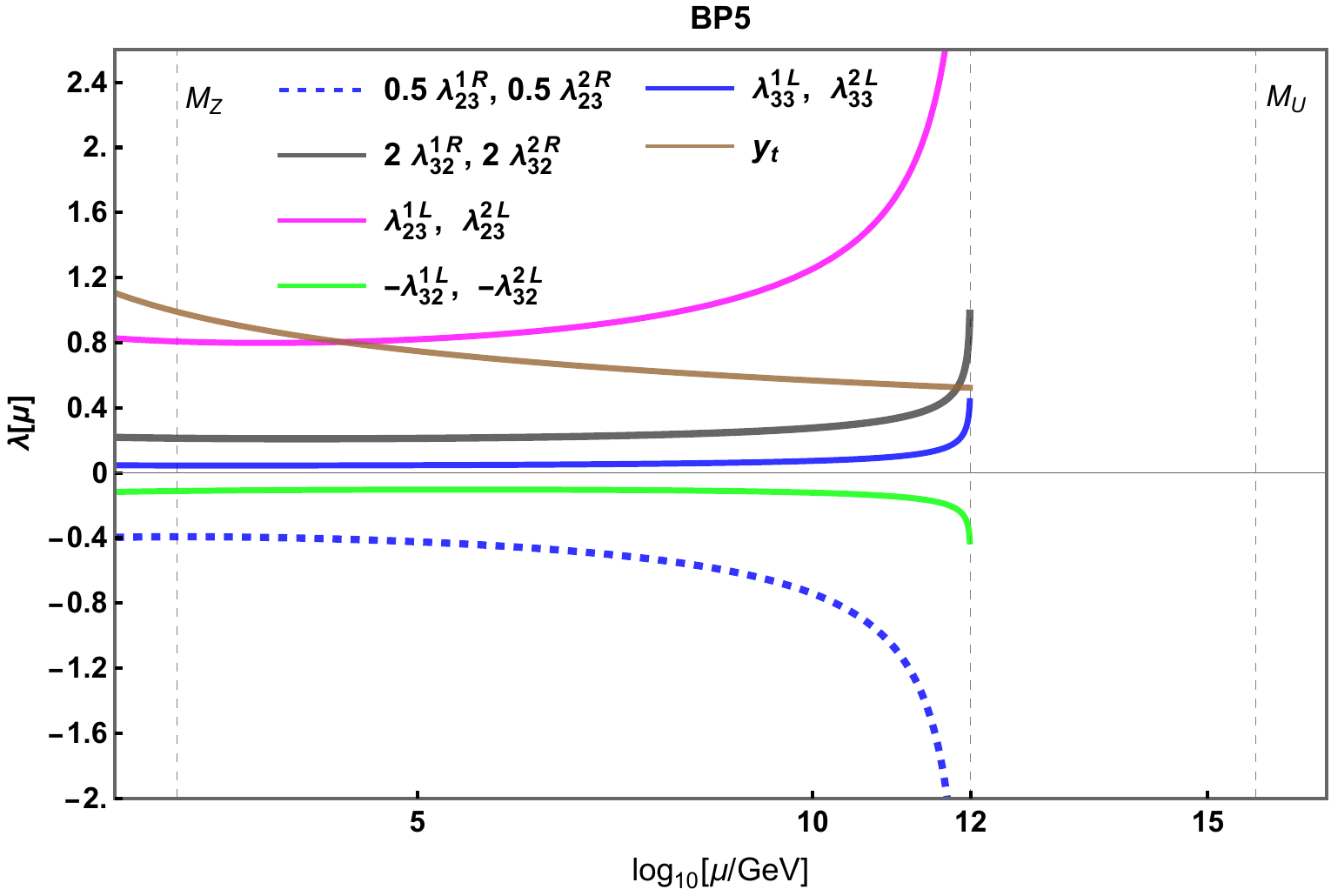}\label{fig:YukawaBP5}}
\caption{The running of the leptoquark and $t$-Yukawa couplings for the 5 selected BPs. The panels (b), (d) and (e) share a common legend, whereas the legend for each of the other panels is shown within it. Some plots have been scaled for clarity.}
\label{fig:Yukawa}
\end{figure*}

\begin{figure*}[tbp]
\captionsetup[subfigure]{labelformat=empty}
\subfloat[(a)]{\includegraphics[scale=0.36]{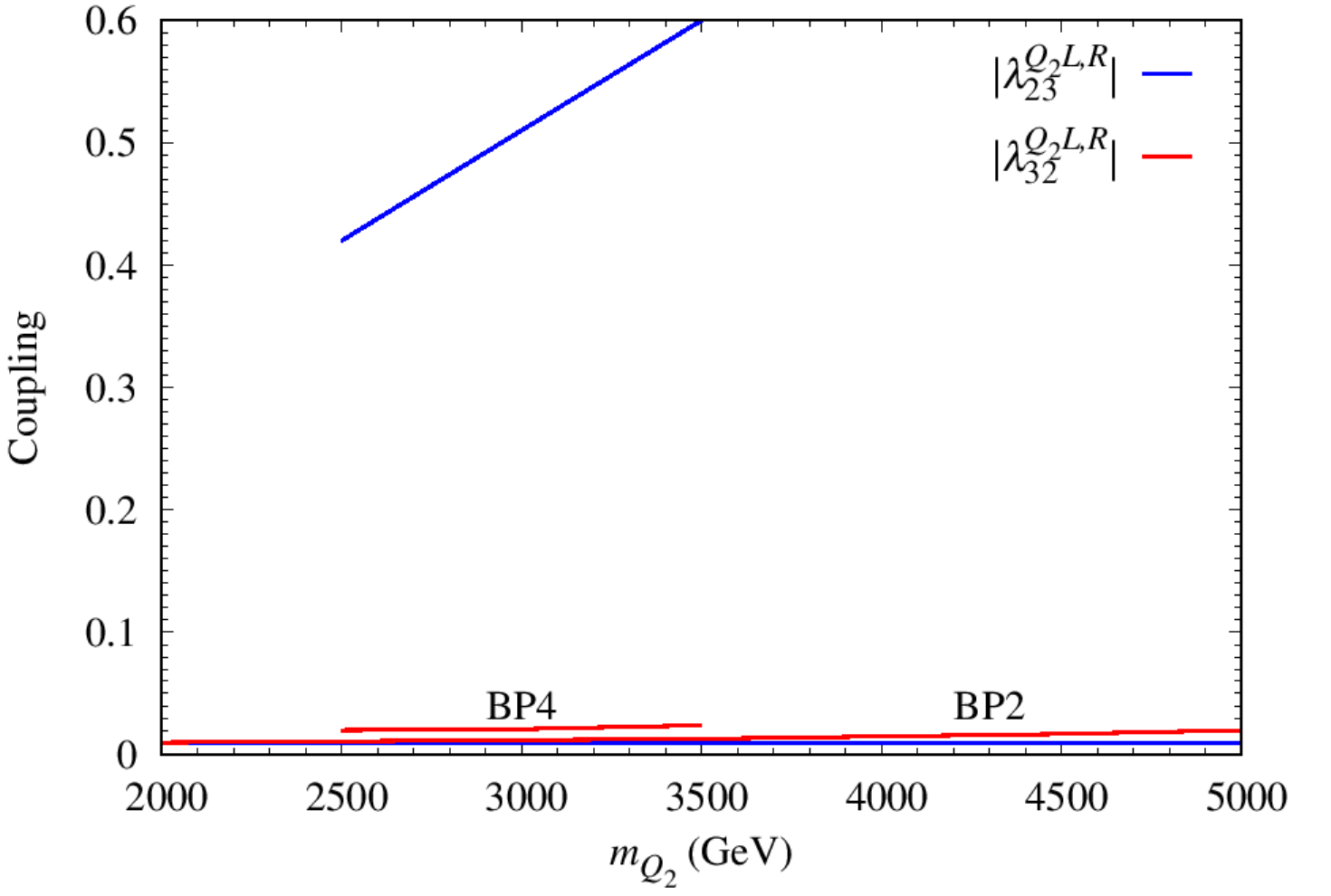}\label{fig:6a}}
\subfloat[(b)]{\hspace*{0.5cm}\includegraphics[scale=0.36]{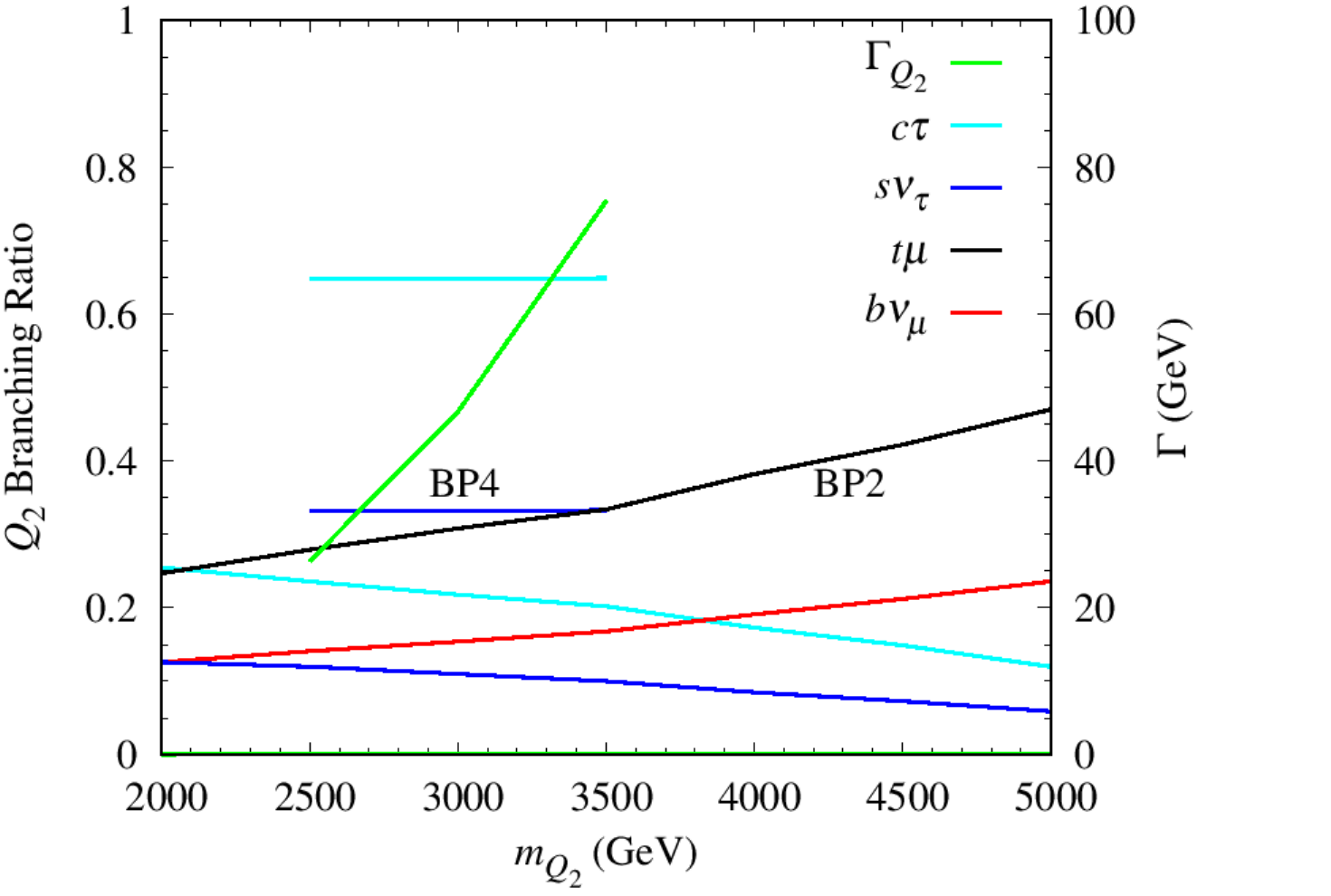}\label{fig:6b}}
\caption{Variation in (a) the $Q_2$ couplings needed to obtain the minimum acceptable values of $R_{D^{(*)}}$ and $\Delta a_\mu$, and (b) its BRs with increasing mass, for BPs 2 (longer lines) and 4 (shorter lines).}
\label{fig:BRs-amu}
\end{figure*}

The RG running of the couplings is illustrated in Fig. \ref{fig:Yukawa} for BPs 1, 2, and 3 in the panels on the left, and for BPs 4 and 5 in the right column. In panels (b), (d) and (e), each of the 11 couplings is shown with a distinct line. In the remaining three panels, corresponding to BPs 1, 3, and 5, each line other than the one for $y_t$ depicts the coupling pairs with identical magnitudes at the EW scale, since the relevant RGEs, given in Appendix \ref{sec:Yukawa}, are symmetric under their interchange. BP1 has the most well-behaved running, as seen in Fig.~\ref{fig:YukawaBP1}, with all the couplings staying in the perturbative realm up to the GUT scale, which is to be expected, since their initial values are fairly small. The running is almost as good in Fig.~\ref{fig:YukawaBP2}, which actually represents a slightly modified BP2, wherein the mass and couplings of $Q_1$ are exactly as in Table~\ref{tab:BPs}, but $m_{Q_2}$ is set to 5\,TeV. This modification is aimed at analysing whether an increase in the mass of $Q_2$ necessitates a large enough rise in the sizes of its couplings (which can be read off from Fig. \ref{fig:6a}) to turn them non-perturbative. But since this BP was selected such that the couplings of $Q_1$ alone are sufficient for satisfying all the requirements, only a very small gradual enhancement in $\lambda^{Q_2L}_{32}=\lambda^{Q_2R}_{32}$ with rising $m_{Q_2}$ is needed for obtaining the desired $\Delta a_\mu$.

For BP3 (Fig.~\ref{fig:YukawaBP3}) with $m_{Q_{1,2}}=2.5$\,TeV, some of the couplings, on account of being initially significantly larger than in BP1, start diverging near the GUT scale, but hit the Landau pole much later. As with BP2 above, the panels (d) and (e) correspond to BP4 with $m_{Q_2}$ increased to 3\,TeV and 3.5\,TeV, respectively, while retaining the original values of the mass (2.5\,TeV) and couplings of $Q_1$. For $m_{Q_2}=3$\,TeV in panel (d), the divergence is much sharper for some of the couplings, compared to BP3, while for $M_{Q_2}=3.5$\,TeV in panel (e), most of the couplings hit the Landau pole just before $M_U$. Finally, due to the significantly larger initial values of some of the couplings in BP5, with $m_{Q_{1,2}}=3$\,TeV, they turn non-perturbative way below the GUT scale, according to Fig. \ref{fig:YukawaBP5}, thus invalidating our framework. Therefore, naively, we can conclude that i) $m_{Q_2}\simeq 3$\,TeV is permissible in our model, as long as $m_{Q_1}$ does not exceed 2.5\,TeV, and ii) none of the $Q_{1,2}$ couplings can have a magnitude $\gtrsim 0.8$ at the EW scale for explaining the experimental data. The individual limits on $\lambda^{Q_{1,2}L}_{23}$ and $\lambda^{Q_{1,2}R}_{23}$ are very close to each other, as one would expect -- the minor difference arises from the unequal couplings of the $Z$ boson to the left- and right-handed fermions. It is, therefore, fair to assume that the limit on the combination $[(\lambda^{Q_{1,2}L}_{23})^2+(\lambda^{Q_{1,2}R}_{23})^2]^{1/2}$ is also very similar.

Fig. \ref{fig:BRs-amu}(a) further illustrates the interplay between the theoretical and experimental constraints, as well as between the mass and couplings of $Q_2$ for BPs 2 and 4. For BP2, as pointed out previously, these couplings remain pertubative even for $m_{Q_2}$ as large as 5\,TeV (and hence more than twice the $m_{Q_1}$), while also fulfilling the conditions listed at the start of this section. In contrast, the large $m_{Q_2}$, and consequently the larger initial values of the ($Q_1$ and) $Q_2$ couplings needed to satisfy these conditions for BP4, drive the model into the non-perturbative realm for $m_{Q_2} \gtrsim 3$\,TeV. Evidently, this interplay thus makes our model rather predictive, unlike many other competing (bottom-up) scenarios that have been investigated in the same context. 

Fig. \ref{fig:BRs-amu}(b) shows some crucial differences in the LHC phenomenology of $Q_2$ between BPs 2 and 4. According to Table \ref{tab:BPs}, the very small $Q_2$ couplings in BP2 lead to its total width being four orders of magnitude smaller than that of $Q_1$. Increasing the $m_{Q_2}$ to 5\,TeV for this BP has no visible impact on $\Gamma_2$ (notice that the corresponding green line for BP2 lies right on top of the bottom axis, and is almost invisible). On the other hand, $m_{Q_{1,2}}=2.5$\,TeV in BP4 already substantially increases the sizes of the couplings of $Q_2$, and hence its total width, which further rises beyond 40\,GeV for $m_{Q_2}\gtrsim 3$\,TeV. Importantly though, $m_{Q_1}=m_{Q_2}$ for both these BPs would appear as a single resonance, with a width that is several tens of GeV, at the LHC. For a mass-splitting of $\mathcal{O}(100)$\,GeV, on the other hand, it might be possible to individually resolve the peaks for $Q_1$ and $Q_2$, which will be mutually well-separated. 

Furthermore, both $Q_1$ and $Q_2$ for BP4 have the highest tree-level decay BR in the $c\tau$ mode, and the second highest one in the $s\nu_\tau$ channel, according to Fig. \ref{fig:BRs-amu}(b). Both these BRs remain almost constant for the entire range of $m_{Q_2}$ (which is however much more strongly restricted by the perturbativity requirement compared to BP2, hence the shorter corresponding lines). For BP2, in contrast, the BRs of $Q_2$ in the $c\tau$ and $t\mu$ decay channels are almost equally dominant for $m_{Q_2}=2$\,TeV, but the latter immediately takes over and is by far the leading mode at $m_{Q_2}=5$\,TeV, where $b\nu_\mu$ is the next dominant channel. These unconventional search channels, despite the sub-GeV total width of $Q_2$, may prove crucial for observing it simultaneously with the $Q_1$ at the LHC. The likelihood of its detection, when the $Q_{1,2}$ couplings are large enough to explain $R_{D^{(*)}}$ and $\Delta a_\mu$ both, but sufficiently small to remain perturbative at the GUT scale, should grow with $m_{Q_2}-m_{Q_1}$.

Finally, in Table \ref{tab:fermionmass}, we give approximate fermion mass values at the GUT scale for the original BPs 1 and 4, as well as for the modified BPs 2 and 4, along with their SM values at the same scale in the second column. The $d$-quark and $e^-$ have the same mass in each of the two cases since they do not get contributions from the $Q_{1,2}$ couplings (see Eqs.~(\ref{Yukawad}) and (\ref{Yukawae})). These GUT values of $m_{d,e}$ are, however, different from the SM in our model, since they run with the modified gauge couplings. The other mass values (and their ratios) change slightly (within $O(1)$) from those interpolated from the SM. Hence, we anticipate the results of the numerical fitting of our model's parameters to the SM data to be very similar to the ones obtained for the $\mathrm{SO}(10)$-based models in a number of previous studies (see Refs.~\cite{Joshipura:2011nn,Meloni:2014rga,Meloni:2016rnt,Altarelli:2013aqa} for recent examples). The purpose of this table is to provide an estimate of the deviation of the masses from the case when we have only the SM at the EW scale, while we leave the full numerical fitting for a future analysis.

% Since there are currently no measurements available form the LHC experiments (regarding the definite values of the new Yukawa couplings) and hence no definite BSM model at the electroweak scale as input, we leave the determination of numerical values of $\mathrm{SO}(10)$ parameters for future work.

\begin{table*}[tbp]
\begin{center}
\begin{tabular}{c||c|c|c|c|c}
%\toprule
\hline
$\vphantom{\Big|}$
%\diaghead{\theadfont Diag ColumnmnHead II}%
  Fermion masses at $M_U$ & $\quad\;\;\;\textrm{SM}\;\;\;\quad$ & $\;\;\;\quad\textrm{BP1}\;\;\;\quad$ & $\;\quad\textrm{BP2}\;(m_{Q_2}=5\,\textrm{TeV})\;\quad$ & $\;\;\;\quad\textrm{BP4}\;\;\;\quad$ & $\;\quad\textrm{BP4}\;(m_{Q_2}=3\,\textrm{TeV})\;\quad$  \\
\hline\hline
$\vphantom{\big|}$
$m_t/{\footnotesize\textrm{GeV}}$  &      $81.13$&                $79.28$     &   $82.03$  &   $82.96$  &   $83.20$ \\
$\vphantom{\big|}$
$m_b/{\footnotesize\textrm{GeV}}$  &      $1.08$&                $1.05$     &   $1.08$ &   $1.09$    &   $1.10$  \\
$\vphantom{\big|}$
$m_c/{\footnotesize\textrm{GeV}}$ 	 & $0.261$&           $0.284$       &     $0.286$   &   $0.317$  &   $0.340$  \\
$\vphantom{\big|}$
$m_s/(10^{-3} {\footnotesize\textrm{GeV}})$ 	 & $23.35$&           $18.40$       &     $18.36$ &   $19.21$ &   $19.7$     \\
$\vphantom{\big|}$
$m_u/(10^{-3} {\footnotesize\textrm{GeV}})$ 	   & $0.482$             & $0.479$     & $0.482$  &   $0.483$  &   $0.483$    \\
$\vphantom{\big|}$
$m_d/(10^{-3} {\footnotesize\textrm{GeV}})$ 	   & $1.229$             & $0.929$     & $ 0.929$  &   $0.929$  &   $0.929$    \\
$\vphantom{\big|}$
$m_{\tau}/{\footnotesize\textrm{GeV}}$ 	  & $1.72$             & $1.75$    &   $1.73$  &   $2.38$ &   $2.93$  \\
$m_{\mu}/(10^{-3} {\footnotesize\textrm{GeV}})$ 	& $101.3$       & $78.7$    & $84.4$  &   $86.5$ &   $87.1$     \\
$\vphantom{\big|}$
$m_{e}/(10^{-3} {\footnotesize\textrm{GeV}})$ 	  & $0.480$             & $0.371$    &   $0.371$  &   $0.371$ &   $0.371$  \\
%$\vphantom{\big|}$
%$\vphantom{\big|}$
%$m_t/m_b$  & $75.24$ & $75.40 $  & $75.39$ &   $75.95$ &   $75.96$ \\
%$\vphantom{\big|}$
%$m_{\tau}/m_b$ 	& $1.60$		& $1.66$ & $1.81$ &   $2.18$ &   $2.68$ \\
%$\vphantom{\big|}$
%$m_{\mu}/m_s$ 		& $4.34$	& $4.27$   & $4.22$ &   $4.50$ &   $4.41$ \\
%$\vphantom{\big|}$
%$m_{e}/m_d$ 	& $0.390$& $0.399$ & $0.399$ &   $0.399$ &   $0.399$  \\
\hline
\end{tabular}
\label{tab:fermionmass}
\end{center}
\caption{Fermion masses at the unification scale $M_U= 10^{15.6}$\,GeV for the original BPs 1 and 4, and for the modified BPs 2 and 4.}
\end{table*}

%%%%%%%%%%%%%%%%%%%%%%%%%%%%%%%%%%%%%%%%%%%%%%%%%%%%%%%%%%%%%%%%%%%%%%%
\section{Conclusions}
\label{sec:Conc}

During recent years, many new physics candidates have been explored in order to explain the $a_\mu$ and $R_{D^{(*)}}$ anomalies that have persisted in the experimental data. One of the strongest candidates that has been claimed to resolve both these anomalies is a scalar leptoquark of the $S_1$ type. Most scenarios invoking such leptoquarks, or other plausible candidates, for this specific purpose, are nevertheless based on a bottom-up approach and do not attempt to offer a UV-complete picture consistent with many other questions pertinent in high energy physics.

In this article, we have taken the top-down approach of linking these anomalies to the $\mathrm{SO}(10)$ GUT group by proposing the existence of an entire complex $\mathbf{10}_H$ multiplet at the TeV scale, circumventing the well-known mass-splitting problem of GUT gauge groups. This multiplet contains two Higgs doublets and two $S_1$-type leptoquarks -- a combination very likely to address the mentioned collider anomalies. The Higgs sector of the resultant low-energy model resembles that of the Type-II 2HDM. 

After discussing in detail the UV-complete framework as well as its low-energy limit -- the Type-II 2HDM augmented with leptoquarks $Q_1$ and $Q_2$ -- we have presented the results of our analysis of some sample configurations of the relevant parameters. This analysis aimed at testing the consistency of the model's predictions with not only the measurements of the $a_\mu$ and $R_{D^{(*)}}$ observables, but also several other important $B$-physics results as well as with the requirement of the perturbativity of the couplings of interest. We have benchmarked two contrasting types of combinations of the $Q_1$ and $Q_2$ couplings, each of which is fairly small to remain perturbative up to the GUT scale without being in conflict with the direct and indirect search results for the $S_1$-type leptoquark from the LHC. 

\section*{Acknowledgements}
We thank Arvind Bhaskar for helping us with the LHC bounds on leptoquark parameters. The work of UA was supported in part by the Chinese Academy of Sciences President's International Fellowship Initiative (PIFI) under Grant No.~2020PM0019; the Institute of High Energy Physics-Beijing, Chinese Academy of Sciences, under Contract No.~Y9291220K (until June 2022); The Scientific and Technological Research Council of T\"urkiye (T\"UB\.ITAK) B\.{I}DEB 2232-A program under project No.~121C067 (from September 2022). SMu would like to acknowledge support from the ICTP through the Associates Programme (2022-2023). 
%\newpage

\appendix

\section{One-loop RG running of gauge couplings with a single intermediate scale}
\label{sec:RG-running}
For a given particle content, the gauge couplings evolve under one-loop RG running in the $\left[M_A,\,M_B\right]$ energy interval as
\begin{eqnarray}
\label{gaugerunning}
\frac{1}{g_{i}^{2}(M_A)} - \dfrac{1}{g_{i}^2(M_B)}
\;=\; \dfrac{a_i}{8 \pi^2}\ln\dfrac{M_B}{M_A}\;,
\end{eqnarray}
where the RG coefficients $a_i$ are given by \cite{Jones:1981we,Lindner:1996tf}
\begin{eqnarray}
\label{1loopgeneral}
a_{i}
= &-&\frac{11}{3}C_{2}(G_i) + \frac{2}{3}\sum_{R_f} T_i(R_f)\cdot d_1(R_f)\cdots d_n(R_f) \cr
&+& \frac{\eta}{3}\sum_{R_s} T_i(R_s)\cdot d_1(R_s)\cdots d_n(R_s)\;.
\end{eqnarray}
The full gauge group is $G=G_i\otimes G_1\otimes...\otimes G_n$. The summation in Eq.~\eqref{1loopgeneral} is over the irreducible representations of chiral fermions ($R_f$) and of scalars ($R_s$) in the second and third terms, respectively. The coefficient $\eta$ is 1 for the complex representation, and 1/2 for the (pseudo-) real one. The symbol $d_j(R)$ denotes the dimension of the representation $R$ under the group $G_{j\neq i}$. Finally, $C_2(G_i)$ represents the quadratic Casimir for the adjoint representation of the group $G_i$, whereas $T_i$ is the Dynkin index of each representation (see Table~\ref{DynkinIndex}). Note that for the $\mathrm{U}(1)$ group, $C_2(G)=0$, and
\begin{equation}
\sum_{f,s}T \;=\; \sum_{f,s}Y^2\;,
\label{U1Dynkin}
\end{equation}
where $Y$ is the $\mathrm{U}(1)_Y$ charge.

%%%%%%%%%%%%%%%%%%%%%%%%%%%%%%%%%%%%%%%%%%%%%%%%%%%%%%%%%%%%%%%%%%%%%%%
\begin{table}[!t]
\begin{center}
{\begin{tabular}{ccccc}
\hline
\ \ \ Representation & $\quad \mathrm{SU}(2)\quad$ & $\quad \mathrm{SU}(3)\quad$ & $\quad \mathrm{SU}(4)\quad$ &\\ 
%\colrule
\hline\hline
$\vphantom{\bigg|}$ 2 &   $\dfrac{1}{2}$ &              $-$ &   $-$ & $\vphantom{\bigg|}$ \\
$\vphantom{\bigg|}$ 3 &                2 &   $\dfrac{1}{2}$ &   $-$ & $\vphantom{\bigg|}$ \\
$\vphantom{\bigg|}$ 4 &                5 &              $-$ &   $\dfrac{1}{2}$ & $\phantom{\bigg|}$ \\ 
$\vphantom{\bigg|}$ 6 &  $\dfrac{35}{2}$ &   $\dfrac{5}{2}$ &   $1$ & $\vphantom{\bigg|}$ \\
$\vphantom{\bigg|}$ 8 &               42 &              $3$ &   $-$ & $\vphantom{\bigg|}$ \\
$\vphantom{\bigg|}$10 & $\dfrac{165}{2}$ &  $\dfrac{15}{2}$ &   $3$ & $\vphantom{\bigg|}$ \\
$\vphantom{\bigg|}$15 &              280 & $10,\dfrac{35}{2}$ &  4 & $\vphantom{\bigg|}$ \\
\hline
\end{tabular}
\label{DynkinIndex}}
\end{center}
\caption{Dynkin indices $T_i$ for various irreducible representations of the $\mathrm{SU}(2)$, $\mathrm{SU}(3)$, and $\mathrm{SU}(4)$ groups. Our normalization convention follows Ref.~\cite{Lindner:1996tf}. Notice that there are two inequivalent 15-dimensional irreducible representations for $\mathrm{SU}(3)$.}
\end{table}

The boundary/matching conditions at the symmetry-breaking scales for the sequence given in Eq.~(\ref{eq:chain}) are the following.
\begin{eqnarray}
M_U & \;:\; & g_L(M_U) \;=\; g_R(M_U) \;=\; g_4(M_U) \;, \vphantom{\bigg|} 
\cr
M_{\mathrm{PS}} & \;:\; & g_3(M_{\mathrm{PS}})=g_4(M_{\mathrm{PS}}) \;,\quad g_2(M_{\mathrm{PS}})\;=\;g_L(M_{\mathrm{PS}})\;,\cr
 & \;\; & \frac{1}{g_1^2(M_{\mathrm{PS}})} \;=\; \frac{1}{g_R^2(M_{\mathrm{PS}})}+ \frac{2}{3}\frac{1}{g_4^2(M_{\mathrm{PS}})}\;,\nn\\
  & \;\; & g_L(M_{\mathrm{PS}})\;=\;g_R(M_{\mathrm{PS}})\;,\\
M_Z & \;:\; & \frac{1}{e^2(M_Z)} \;=\; \frac{1}{g_1^2(M_Z)}+\frac{1}{g_2^2(M_Z)}\;. \nonumber
\label{Matching}
\end{eqnarray}

Together with the matching and boundary conditions given in Eq.~(\ref{Matching}), the one-loop RG running leads to the following conditions on the $M_U$ and $M_{\mathrm{PS}}$ scales.
\begin{eqnarray}
\label{unification-relations}
2\pi\left[\dfrac{3-8\sin^2\theta_W(M_Z)}{\alpha(M_Z)}\right]
 = 
 \left(3a_1 -5a_2\right)\ln\dfrac{M_{\mathrm{PS}}}{M_Z}&&\nn\\
\hspace{-.75cm}+\left(-5a_L+3a_R+2a_4\right)\ln\dfrac{M_U}{M_{\mathrm{PS}}}
\;,&&
%\vphantom{\Bigg|}
\cr
%%%%%
2\pi\left[\dfrac{3}{\alpha(M_Z)} - \dfrac{8}{\alpha_s(M_Z)}\right]
 = 
 \left(3a_1 + 3a_2 - 8a_3\right)\ln\dfrac{M_{\mathrm{PS}}}{M_Z}&&\nn\\
+\left(3a_L+3a_R-6a_4\right)\ln\dfrac{M_U}{M_{\mathrm{PS}}}
\;,\hspace{0.5cm}&&
%\vphantom{\Bigg|}
%\cr
%& &\hspace{-0.75cm}~
\end{eqnarray}
where the notation of $a_i$ is self-evident. The unified gauge coupling $\alpha_U$ at the scale $M_U$ is then found as
\begin{eqnarray}
\label{A6}
\dfrac{2\pi}{\alpha_U}
 =  \dfrac{2\pi}{\alpha_s(M_Z)}
-\left( a_4\;\ln\dfrac{M_U}{M_{\mathrm{PS}}}
+ a_3\;\ln\dfrac{M_{\mathrm{PS}}}{M_Z}
\right)
\;.
\end{eqnarray}
%
%%
%%%%%%%%%%%%%%%%%%%%%%%%%%%%%%%%%%%%%%%%%%%%%%%%%%%%%%%%%%%%%
\section{One-loop RG equations of the Yukawa couplings}
\label{sec:RGEs}

The new Yukawa matrices in the Lagrangian given in Eq.~\eqref{eq:lagrangianLQ} are defined in our setup as
\begin{eqnarray}
\mathbf{\Lambda}^{1L} &\rightarrow& \left(\begin{array}{ccc}0 & 0 & 0 \\0 & 0 & \lambda^{Q_1 L}_{23} \\0 & \lambda^{Q_1 L}_{32} & \lambda^{Q_1 L}_{33} \end{array}\right),\;\; \mathbf{\Lambda}^{1R} \rightarrow \left(\begin{array}{ccc}0 & 0 & 0 \\0 & 0 & \lambda^{Q_1 R}_{23} \\0 & \lambda^{Q_1 R}_{32} &0 \end{array}\right)\qquad \nonumber\\
\mathbf{\Lambda}^{2L} &\rightarrow& \left(\begin{array}{ccc}0 & 0 & 0 \\0 & 0 & \lambda^{Q_2 L}_{23} \\0 & \lambda^{Q_2 L}_{32} & \lambda^{Q_2 L}_{33} \end{array}\right),\;\; \mathbf{\Lambda}^{2R} \rightarrow \left(\begin{array}{ccc}0 & 0 & 0 \\0 & 0 & \lambda^{Q_2 R}_{23} \\0 & \lambda^{Q_2 R}_{32} &0 \end{array}\right).\nonumber
\end{eqnarray}

The corresponding 1-loop RG equations, following Ref.~\cite{Machacek:1983fi} (or, alternatively, obtained by implementing the model in SARAH), are given below. 

\begin{eqnarray}
&&16 \pi^2 \beta_{\lambda^{Q_2 L}_{33}}=4\left(\lambda_{33}^{Q_2 L}\right)^3+\lambda_{33}^{Q_2 L}\left[-\frac{5}{6} g_1^2-\frac{9}{2} g_2^2-4 g_3^2 \right.\nonumber\\
&&+\frac{y_t^2}{2}+\frac{3}{2}\left(\lambda_{23}^{Q_1 L}\right)^2+\frac{1}{2}\left(\lambda_{32}^{Q_1 L}\right)^2+4\left(\lambda_{33}^{Q_1 L}\right)^2\nonumber\\
&&+4\left(\lambda_{23}^{Q_2 L}\right)^2\left.+4\left(\lambda_{32}^{Q_2 L}\right)^2+\left(\lambda_{23}^{Q_2 R}\right)^2+\left(\lambda_{32}^{Q_2 R}\right)^2\right]\nonumber\\
&&+\lambda_{33}^{Q_1 L} \left[\frac{5}{2} \lambda_{23}^{Q_1 L} \lambda_{23}^{Q_2 L}+\frac{7}{2} \lambda_{32}^{Q_1 L} \lambda_{32}^{Q_2 L}\right.\nonumber\\
&&\left.+\lambda_{23}^{Q_1 R} \lambda_{23}^{Q_2 R}+\lambda_{32}^{Q_1 R} \lambda_{32}^{Q_2 R}\right]\;,
\end{eqnarray}

%%%%%%%%%%%%%%%%%%%%
\begin{eqnarray}
&&16 \pi^2 \beta_{\lambda^{Q_2 L}_{32}}=4\left(\lambda_{32}^{Q_2 L}\right)^3+\lambda_{32}^{Q_2 L}\left[-\frac{5}{6} g_1^2-\frac{9}{2} g_2^2-4 g_3^2 \right.\nonumber\\
&&+\frac{y_t^2}{2}+4\left(\lambda_{32}^{Q_1 L}\right)^2+\frac{1}{2}\left(\lambda_{33}^{Q_1 L}\right)^2+2\left(\lambda_{23}^{Q_2 L}\right)^2
\nonumber\\
&&\left.+4\left(\lambda_{33}^{Q_2 L}\right)^2+\left(\lambda_{23}^{Q_2 R}\right)^2+\left(\lambda_{32}^{Q_2 R}\right)^2\right]\nonumber\\
&&+\lambda_{32}^{Q_1 L} \left[2 \lambda_{23}^{Q_1 L} \lambda_{23}^{Q_2 L}+\frac{7}{2} \lambda_{33}^{Q_1 L} \lambda_{33}^{Q_2 L}\right.\nonumber\\
&&\left.+\lambda_{23}^{Q_1 R} \lambda_{23}^{Q_2 R}+\lambda_{32}^{Q_1 R} \lambda_{32}^{Q_2 R}\right]\;,
\end{eqnarray}
%%%%%%%%%%%%%%%%%%%%
\begin{eqnarray}
&&16 \pi^2 \beta_{\lambda^{Q_2 L}_{23}}=4\left(\lambda_{23}^{Q_2 L}\right)^3+\lambda_{23}^{Q_2 L}\left[-\frac{5}{6} g_1^2-\frac{9}{2} g_2^2-4 g_3^2 \right.\nonumber\\
&&+4\left(\lambda_{23}^{Q_1 L}\right)^2+\frac{3}{2}\left(\lambda_{33}^{Q_1 L}\right)^2+2\left(\lambda_{32}^{Q_2 L}\right)^2 \nonumber\\
&&\left.+4\left(\lambda_{33}^{Q_2 L}\right)^2+\left(\lambda_{23}^{Q_2 R}\right)^2+\left(\lambda_{32}^{Q_2 R}\right)^2\right]\nonumber\\
&&+\lambda_{23}^{Q_1 L} \left[2 \lambda_{32}^{Q_1 L} \lambda_{32}^{Q_2 L}+\frac{5}{2} \lambda_{33}^{Q_1 L} \lambda_{33}^{Q_2 L}\right.\nonumber\\
&&\left.+\lambda_{23}^{Q_1 R} \lambda_{23}^{Q_2 R}+\lambda_{32}^{Q_1 R} \lambda_{32}^{Q_2 R}\right]\;,
\end{eqnarray}
%%%%%%%%%%%%%%%%%%%%
\begin{eqnarray}
&&16 \pi^2 \beta_{\lambda^{Q_1 L}_{33}}=4\left(\lambda_{33}^{Q_1 L}\right)^3+\lambda_{33}^{Q_1 L}\left[-\frac{5}{6} g_1^2-\frac{9}{2} g_2^2-4 g_3^2 \right.\nonumber\\
&&+\frac{y_t^2}{2}+4\left(\lambda_{23}^{Q_1 L}\right)^2+4\left(\lambda_{32}^{Q_1 L}\right)^2+\frac{3}{2}\left(\lambda_{23}^{Q_2 L}\right)^2\nonumber\\
&&\left.+\frac{1}{2}\left(\lambda_{32}^{Q_2 L}\right)^2+4\left(\lambda_{33}^{Q_2 L}\right)^2+\left(\lambda_{23}^{Q_1 R}\right)^2+\left(\lambda_{32}^{Q_1 R}\right)^2\right]\nonumber\\
&&+\lambda_{33}^{Q_2 L} \left[\frac{5}{2} \lambda_{23}^{Q_1 L} \lambda_{23}^{Q_2 L}+\frac{7}{2} \lambda_{32}^{Q_1 L} \lambda_{32}^{Q_2 L}\right.\nonumber\\
&&\left.+\lambda_{23}^{Q_1 R} \lambda_{23}^{Q_2 R}+\lambda_{32}^{Q_1 R} \lambda_{32}^{Q_2 R}\right]\;,
\end{eqnarray}

\begin{eqnarray}
&&16 \pi^2 \beta_{\lambda^{Q_1 L}_{32}}=4\left(\lambda_{32}^{Q_1 L}\right)^3+\lambda_{32}^{Q_1 L}\left[-\frac{5}{6} g_1^2-\frac{9}{2} g_2^2-4 g_3^2\right.\nonumber\\
&&+\frac{y_t^2}{2} +4\left(\lambda_{33}^{Q_1 L}\right)^2+2\left(\lambda_{23}^{Q_1 L}\right)^2+\left(\lambda_{23}^{Q_1 R}\right)^2\nonumber\\
&&\left.+\left(\lambda_{32}^{Q_1 R}\right)^2+4\left(\lambda_{32}^{Q_2 L}\right)^2+\frac{1}{2}\left(\lambda_{33}^{Q_1 L}\right)^2\right]\nonumber\\
&&+\lambda_{32}^{Q_2 L} \left[2 \lambda_{23}^{Q_1 L} \lambda_{23}^{Q_2 L}+\frac{7}{2} \lambda_{33}^{Q_1 L} \lambda_{33}^{Q_2 L}\right.\nonumber\\
&&\left.+\lambda_{23}^{Q_1 R} \lambda_{23}^{Q_2 R}+\lambda_{32}^{Q_1 R} \lambda_{32}^{Q_2 R}\right]\;,
\end{eqnarray}
%%%%%%%%%%%%%%%%%%%%%%
\begin{eqnarray}
&&16 \pi^2 \beta_{\lambda^{Q_1 L}_{23}}=4\left(\lambda_{23}^{Q_1 L}\right)^3+\lambda_{23}^{Q_1 L}\left[-\frac{5}{6} g_1^2-\frac{9}{2} g_2^2-4 g_3^2 \right.\nonumber\\
&&+2\left(\lambda_{32}^{Q_1 L}\right)^2+4\left(\lambda_{33}^{Q_1 L}\right)^2+4\left(\lambda_{23}^{Q_2 L}\right)^2\nonumber\\
&&\left.+\frac{3}{2}\left(\lambda_{33}^{Q_2 L}\right)^2+\left(\lambda_{23}^{Q_1 R}\right)^2+\left(\lambda_{32}^{Q_1 R}\right)^2\right]\nonumber\\
&&+\lambda_{23}^{Q_2 L} \left[2 \lambda_{32}^{Q_1 L} \lambda_{32}^{Q_2 L}+\frac{5}{2} \lambda_{33}^{Q_1 L} \lambda_{33}^{Q_2 L}\right.\nonumber\\
&&\left.+\lambda_{23}^{Q_1 R} \lambda_{23}^{Q_2 R}+\lambda_{32}^{Q_1 R} \lambda_{32}^{Q_2 R}\right]\;,
\end{eqnarray}
%%%%%%%%%%%%%%%%%%%%%%
\begin{eqnarray}
&&16 \pi^2 \beta_{\lambda^{Q_2 R}_{32}}=3\left(\lambda_{32}^{Q_2 R}\right)^3+\lambda_{32}^{Q_2 R}\left[-\frac{13}{3} g_1^2-4 g_3^2 +y_t^2\right.\nonumber\\
&&\left.+2\left(\lambda_{23}^{Q_2 L}\right)^2+2\left(\lambda_{32}^{Q_2 L}\right)^2+2\left(\lambda_{33}^{Q_2 L}\right)^2\right]\nonumber\\
&&+\left(\lambda_{23}^{Q_2 R}\right)^2+3\left(\lambda_{32}^{Q_1 R}\right)^2+2\lambda_{32}^{Q_1 R} \left[ \lambda_{23}^{Q_1 L} \lambda_{23}^{Q_2 L}\right.\nonumber\\
&&\left.+\lambda_{32}^{Q_1 L} \lambda_{32}^{Q_2 L}+ \lambda_{33}^{Q_1 L} \lambda_{33}^{Q_2 L}+\frac{1}{2} \lambda_{23}^{Q_1 R}\lambda_{23}^{Q_2 R}\right]\;,
\end{eqnarray}
%%%%%%%%%%%%%%%%%%%%%%
\begin{eqnarray}
&&16 \pi^2 \beta_{\lambda^{Q_2 R}_{23}}=3\left(\lambda_{23}^{Q_2 R}\right)^3+\lambda_{23}^{Q_2 R}\left[-\frac{13}{3} g_1^2-4 g_3^2 \right.\nonumber\\
&&\left.+2\left(\lambda_{23}^{Q_2 L}\right)^2+2\left(\lambda_{32}^{Q_2 L}\right)^2+2\left(\lambda_{33}^{Q_2 L}\right)^2+\left(\lambda_{32}^{Q_2 R}\right)^2\right.\nonumber\\
&&\left.+3\left(\lambda_{23}^{Q_1 R}\right)^2\right]+2\lambda_{23}^{Q_1 R} \left[ \lambda_{23}^{Q_1 L} \lambda_{23}^{Q_2 L}+\lambda_{32}^{Q_1 L} \lambda_{32}^{Q_2 L}\right.\nonumber\\
&&\left.+ \lambda_{33}^{Q_1 L} \lambda_{33}^{Q_2 L}+\frac{1}{2} \lambda_{32}^{Q_1 R}\lambda_{32}^{Q_2 R}\right]\;,
\end{eqnarray}
%%%%%%%%%%%%%%%%%%%%%%
\begin{eqnarray}
&&16 \pi^2 \beta_{\lambda^{Q_1 R}_{32}}=3\left(\lambda_{32}^{Q_1 R}\right)^3+\lambda_{32}^{Q_1 R}\left[-\frac{13}{3} g_1^2-4 g_3^2 +y_t^2\right.\nonumber\\
&&\left.+2\left(\lambda_{23}^{Q_1 L}\right)^2+2\left(\lambda_{32}^{Q_1 L}\right)^2+2\left(\lambda_{33}^{Q_1 L}\right)^2+\left(\lambda_{23}^{Q_1 R}\right)^2\right.\nonumber\\
&&\left.+3\left(\lambda_{32}^{Q_2 R}\right)^2\right]+2\lambda_{32}^{Q_2 R} \left[ \lambda_{23}^{Q_1 L} \lambda_{23}^{Q_2 L}+\lambda_{32}^{Q_1 L} \lambda_{32}^{Q_2 L}\right.\nonumber\\
&&\left.+ \lambda_{33}^{Q_1 L} \lambda_{33}^{Q_2 L}+\frac{1}{2} \lambda_{23}^{Q_1 R}\lambda_{23}^{Q_2 R}\right]\;,
\end{eqnarray}

\begin{eqnarray}
&&16 \pi^2 \beta_{\lambda^{Q_1 R}_{23}}=3\left(\lambda_{23}^{Q_1 R}\right)^3+\lambda_{23}^{Q_1 R}\left[-\frac{13}{3} g_1^2-4 g_3^2 \right.\nonumber\\
&&\left.+2\left(\lambda_{23}^{Q_1 L}\right)^2+2\left(\lambda_{32}^{Q_1 L}\right)^2+2\left(\lambda_{33}^{Q_1 L}\right)^2+\left(\lambda_{32}^{Q_1 R}\right)^2\right.\nonumber\\
&&\left.+3\left(\lambda_{23}^{Q_2 R}\right)^2\right]+2\lambda_{23}^{Q_2 R} \left[ \lambda_{23}^{Q_1 L} \lambda_{23}^{Q_2 L}+\lambda_{32}^{Q_1 L} \lambda_{32}^{Q_2 L}\right.\nonumber\\
&&\left.+ \lambda_{33}^{Q_1 L} \lambda_{33}^{Q_2 L}+\frac{1}{2} \lambda_{32}^{Q_1 R}\lambda_{32}^{Q_2 R}\right]\;,
\end{eqnarray}
%%%%%%%%%%%%%%%%%%%%%%
\begin{eqnarray}
&&16 \pi^2 \beta_{y_t}=\frac{9}{2}y_t^3+\frac{y_t}{2}\left[-\frac{17}{6}g_1^2-\frac{9}{2}g_2^2-16g_3^2\quad\quad\quad\quad\right.\nonumber\\
&&\left.+\left(\lambda_{33}^{Q_1 L}\right)^2+\left(\lambda_{33}^{Q_2 L}\right)^2+\left(\lambda_{32}^{Q_1 L}\right)^2+\left(\lambda_{32}^{Q_2 L}\right)^2\right.\nonumber\\
&&\left.+ \left(\lambda_{32}^{Q_1 R}\right)^2+ \left(\lambda_{32}^{Q_2 R}\right)^2\right]\;,
\end{eqnarray}
%%%%%%%%%%%%%%%%%%%%%%
\begin{eqnarray}
&&16 \pi^2 \beta_{y_c}=\frac{y_c}{2}\left[6y_t^2-\frac{17}{6}g_1^2-\frac{9}{2}g_2^2-16 g_3^2+\left(\lambda_{23}^{Q_1 L}\right)^2\right.\nonumber\\
&&\left.+\left(\lambda_{23}^{Q_2 L}\right)^2+ \left(\lambda_{23}^{Q_1 R}\right)^2+ \left(\lambda_{23}^{Q_2 R}\right)^2\right]\;,
\end{eqnarray}
%%%%%%%%%%%%%%%%%%%%%%
\begin{eqnarray}
16 \pi^2 \beta_{y_u}=\frac{y_u}{2} \left[6y_t^2-\frac{17}{6}g_1^2-\frac{9}{2}g_2^2-16 g_3^2\right]\;,\quad
\end{eqnarray}
%%%%%%%%%%%%%%%%%%%%%%
\begin{eqnarray}
&&16 \pi^2 \beta_{y_b}=\frac{y_b}{2}\left[3y_t^2-\frac{5}{6}g_1^2-\frac{9}{2}g_2^2-16 g_3^2+\left(\lambda_{32}^{Q_1 L}\right)^2\right.\nonumber\\
&&\left.+\left(\lambda_{32}^{Q_2 L}\right)^2+ \left(\lambda_{33}^{Q_1 L}\right)^2+ \left(\lambda_{33}^{Q_2 L}\right)^2\right]\;,
\end{eqnarray}
%%%%%%%%%%%%%%%%%%%%%%
\begin{eqnarray}
&&16 \pi^2 \beta_{y_s}=\frac{y_s}{2}\left[-\frac{5}{6}g_1^2-\frac{9}{2}g_2^2-16 g_3^2\quad\quad\quad\quad\quad\quad\quad \right.\nonumber\\
&&\left.+\left(\lambda_{23}^{Q_1 L}\right)^2+\left(\lambda_{23}^{Q_2 L}\right)^2\right]\;,
\end{eqnarray}
%%%%%%%%%%%%%%%%%%%%%%
\begin{eqnarray}
\label{Yukawad}16 \pi^2 \beta_{y_d}=\frac{y_d}{2} \left[-\frac{5}{6}g_1^2-\frac{9}{2}g_2^2-16 g_3^2\right]\;,\quad\quad\quad\;
\end{eqnarray}
%%%%%%%%%%%%%%%%%%%%%%
\begin{eqnarray}
\label{Yukawae}16 \pi^2 \beta_{y_e}=\frac{y_e}{2} \left[-\frac{15}{2}g_1^2-\frac{9}{2}g_2^2\right]\;,\quad\quad\quad\quad\quad\quad
\end{eqnarray}
\\
%%%%%%%%%%%%%%%%%%%%%%
\begin{eqnarray}
&&16 \pi^2 \beta_{y_\tau}=\frac{3y_\tau}{2}\left[-\frac{5}{2}g_1^2-\frac{3}{2}g_2^2+\left(\lambda_{23}^{Q_1 L}\right)^2\quad\quad\quad\quad\right.\nonumber\\
&&\left.+\left(\lambda_{23}^{Q_2 L}\right)^2+\left(\lambda_{33}^{Q_1 L}\right)^2+\left(\lambda_{33}^{Q_2 L}\right)^2\right.\nonumber\\
&&\left. +\left(\lambda_{23}^{Q_1 R}\right)^2+ \left(\lambda_{23}^{Q_2 R}\right)^2\right]\;,
\end{eqnarray}
%%%%%%%%%%%%%%%%%%%%%%
\begin{eqnarray}
&&16 \pi^2 \beta_{y_\mu}=\frac{3 y_\mu}{2}\left[-\frac{5}{2}g_1^2-\frac{3}{2}g_2^2+\left(\lambda_{32}^{Q_1 L}\right)^2\quad\quad\quad\quad\right.\nonumber\\
&&\left.+\left(\lambda_{32}^{Q_2 L}\right)^2+ \left(\lambda_{32}^{Q_1 R}\right)^2+ \left(\lambda_{32}^{Q_2 R}\right)^2\right]\;.
\end{eqnarray}

%\newpage
%\bibliography{2HDM-LQ}{}
%\bibliographystyle{JHEPCust}


\end{document}